\documentclass[sigconf]{acmart}

\AtBeginDocument{%
  }

\copyrightyear{2024} 
\acmYear{2024} 
\setcopyright{acmlicensed}\acmConference[RecSys '24]{18th ACM Conference on Recommender Systems}{October 14--18, 2024}{Bari, Italy}
\acmBooktitle{18th ACM Conference on Recommender Systems (RecSys '24), October 14--18, 2024, Bari, Italy}
\acmDOI{10.1145/3640457.3688107}
\acmISBN{979-8-4007-0505-2/24/10}

\usepackage{arydshln}
\usepackage{multirow}
\usepackage{epstopdf} 
\usepackage{multirow}
\usepackage{subfigure}
\usepackage{tabularx}
\usepackage{hyperref}
\begin{document}

\title{The Elephant in the Room: Rethinking the Usage of Pre-trained Language Model in Sequential Recommendation}

\author{Zekai Qu}
\authornote{Both authors have equal contributions. Ruobing Xie is the corresponding author.}
\affiliation{\institution{China University of Geosciences Beijing}
\city{Beijing}
\country{China}}
\email{zekai_qu@163.com}

\author{Ruobing Xie}
\authornotemark[1]
\affiliation{\institution{Tencent Inc.}
\city{Beijing}
\country{China}}
\email{xrbsnowing@163.com}

\author{Chaojun Xiao}
\affiliation{\institution{Tsinghua University}
\city{Beijing}
\country{China}}
\email{xcjthu@gmail.com}

\author{Xingwu Sun}
\affiliation{\institution{Tencent Inc.}
\city{Beijing}
\country{China}}
\email{sammsun@tencent.com}

\author{Zhanhui Kang}
\affiliation{\institution{Tencent Inc.}
\city{Shenzhen}
\country{China}}
\email{kegokang@tencent.com}


\begin{abstract}
Sequential recommendation (SR) has seen significant advancements with the help of Pre-trained Language Models (PLMs). Some PLM-based SR models directly use PLM to encode user historical behavior's text sequences to learn user representations, while there is seldom an in-depth exploration of the capability and suitability of PLM in behavior sequence modeling. In this work, we first conduct extensive model analyses between PLMs and PLM-based SR models, discovering great underutilization and parameter redundancy of PLMs in behavior sequence modeling. Inspired by this, we explore different lightweight usages of PLMs in SR, aiming to maximally stimulate the ability of PLMs for SR while satisfying the efficiency and usability demands of practical systems. We discover that adopting behavior-tuned PLMs for item initializations of conventional ID-based SR models is the most economical framework of PLM-based SR, which would not bring in any additional inference cost but could achieve a dramatic performance boost compared with the original version. Extensive experiments on five datasets show that our simple and universal framework leads to significant improvement compared to classical SR and SOTA PLM-based SR models without additional inference costs. Our code can be found in \hyperref[https://github.com/777pomingzi/Rethinking-PLM-in-RS]{https://github.com/777pomingzi/Rethinking-PLM-in-RS}.
\end{abstract}



\maketitle

\section{Introduction}

Recommender system is an information processing system that is deployed to forecast user preferences and provide suitable items or information. Its application spans various domains such as e-commerce, advertising, and streaming services. Intensive studies in various real-world scenarios have found that historical interactions are important signals to model user preferences. Hence, sequential recommendation (SR) is a promising topic in the recommender systems community, aiming to capture the users' dynamic preferences and predict the next item they like based on their historical behaviors. Early studies mainly adopt Markov chains to model the transition patterns of users \cite {FPMC}. In recent years, due to the rapid development of deep learning, different kinds of neural networks have been introduced into the SR, which resulted in notable enhancements. The most representative works include GRU4Rec \cite{GRURec}, Caser \cite{Tang_Wang_2018}, SR-GNN \cite{SR-GNN}, SASRec \cite{SASRec}, and BERT4Rec \cite{BERT4Rec}.

With the thriving of PLMs, researchers begin to explore their application in SR. Some studies utilize PLMs as item text encoders, aiming to enhance item representations with the rich prior knowledge embedded within PLMs \cite{zhou2020s3,IDA-SR,TransRec,UniRec,yuan2023go}. These methods always leverage a PLM to encode item text (e.g., title, brand) into text embeddings, which are then utilized to substitute or augment the original ID embeddings. Recently, propelled by the significant reasoning and long text modeling abilities demonstrated by Large Language Models (LLMs), some pioneering researchers have sought to harness the advanced sequence modeling capabilities of PLMs to enhance the performance of SR. One line of work uses fixed LLMs for recommendations via prompt or in-context learning, which achieve promising performance under the few-shot or zero-shot settings \cite{gao2023chat,liu2023chatgpt}. However, due to the significant disparity between language and behavior modeling, these methods often perform poorly when interactions are sufficient. In that case, recent works choose to adopt recommendation objectives to fine-tune PLMs for adaptation, which achieve promising results \cite{geng2022recommendation,cui2022m6,zhang2023recommendation,li2023text,qu2023thoroughly}.

Despite the efficacy of PLM-enhanced methods in existing literature, a critical question remains largely unexplored: do PLMs effectively and efficiently boost sequential recommendation by providing accurate prior knowledge and superior sequence modeling capabilities as theoretically expected? While some research has investigated their effect in item representation learning \cite{TransRec,yuan2023go,wei2023llmrec}, their impact on behavior sequence modeling has yet to be thoroughly examined. Consequently, our study aims to figure out two questions: (1) whether the powerful sequence modeling capabilities of PLMs have been fully and economically utilized in the \emph{behavior sequence modeling} of SR. (2) If not, is there a more \emph{effective, universal, and economical} way for us to employ PLMs for SR?

For question one, we conduct two motivating experiments (in Sec. \ref{analysis}) using a representative SOTA PLM-enhanced multi-domain SR model, RECFORMER \cite{li2023text}. The findings highlight two critical aspects: (1) RECFORMER functions differently from its base PLM backbone in behavior modeling. Its attention maps exhibit a clear functional stratification, with higher layers (more crucial for SR adaptation) more assimilated to conventional ID-based SR models. (2) There appears to be significant parameter redundancy when employing PLMs for behavior modeling. The global attentions of the sequence display considerable resemblance among different heads and layers in the same functional stratification. Only tuning 1/4 selected layers of RECFORMER can yield performance comparable to or even surpassing the original RECFORMER.

Based on the above observation, we implement several enhanced variants of RECFORMER with simplified sequence modeling methods borrowed from classical ID-based SR models such as SASRec and BERT4Rec. We discover the key factor of improvement borrowed from PLM to SR: \textbf{item representations built by behavior-tuned PLM}.
Experimental results indicate that PLM-involved SR models with simpler sequence models can achieve comparable results to RECFORMER with much better efficiency, suggesting that PLMs are not that suitable for behavior sequence modeling, for their superior sequence modeling and reasoning capacities in NLP are not fully leveraged and transferred into the task of SR. Consequently, we conclude a simple, effective, and universal framework to efficiently take advantage of PLM for SR: (a) \textbf{adopting simplified (ID-based) sequential models for behavior sequence modeling} (for question one), and (b) \textbf{using behavior-tuned PLMs (rather than vanilla PLMs) for item embedding initialization} (for question two). Further experiments demonstrate that our behavior-tuned PLMs initialized item embeddings are universal with different (ID-based or PLM-based) sequential models and settings.
The contributions are concluded as follows:

\begin{itemize}

    \item To the best of our knowledge, we are the first work to conduct a thorough analysis of the mechanism and effectiveness of PLMs in behavior sequence modeling of SR. 
    \item We confirm the existence of functional stratification and parameter redundancy in the behavior sequence modeling processes of the current PLM-based SR model.
    \item  We introduce a novel framework that uses behavior-tuned PLMs for item initialization and simplified methods for behavior sequence modeling. Our framework is simple, effective, and universal, leading to significant improvements over PLM-based and classical ID-based SR baselines on different settings without additional inference costs.
\end{itemize}

\section{Related Works}
\subsection{Sequential Recommendation}
Sequential recommendation (SR) aims to model the users' dynamic behavior patterns based on their historical interactions. The very early works usually adopt Markov chains (MCs) to model the behavioral patterns of users. \citeauthor{FPMC} \cite{FPMC} integrates MF and MCs to simultaneously model users' overarching preferences and their historical interactions. Beyond first-order MCs, several approaches have embraced higher-order MCs, incorporating more preceding items for sequential modeling \cite{He_Kang_McAuley_2017,Ruining_Julian_2016}. Later, with the development of deep learning methods, various deep neural networks were introduced into SR. RNNs such as Gated Recurrent Unit (GRU) \cite{GRU} are introduced to model sequential behaviors, including session-based GRU (e.g., GRU4Rec \cite{GRURec}) and user-based GRU \cite{Donkers_Loepp_Ziegler_2017}). Moreover, CNNs have also been demonstrated to effectively capture short-term sequential dynamics using both horizontal and vertical convolutional filters \cite{Tang_Wang_2018}. Many Graph Neural Network (GNN) based models \cite{SR-GNN,gnn2022evolutionary,gnn2022exploiting,gnnsr}, were also implemented by researchers to seek better performance. Recently, due to the notable achievements in sequence modeling made by Transformers \cite{transformers} in natural language processing, many attention-based models have also been intensively explored. SASRec \cite{SASRec} and BERT4Rec \cite{BERT4Rec} employ unidirectional and bidirectional Transformers separately for learning dynamic user preferences. LSSA \cite{LSSA} combines long-term and short-term self-attention mechanisms, designed to model users' enduring preferences alongside their immediate needs. Self-supervised learning (SSL) is also introduced by recent works for improving the performance and training efficiency of SR models \cite{qiu2022contrastive,chen2022intent,xie2022contrastive,wu2022selective,zhou2020s3}. Besides, some recent efforts explore the cross-domain sequential recommendation \cite{ma2019pi,hao2021adversarial,cao2022contrastive}, while most of them face challenges in smoothly transferring the generalized information to new domains.  To address this, some works \cite{TransRec,UniRec,yuan2023go} further use PLMs and vision encoders to fuse modality information into item representations for a more comprehensive understanding of item contents, functioning as the bridge across different datasets.

\subsection{PLM for Recommendation}

With the thriving of foundation models, many applications of PLMs for SR have been researched, hoping to leverage the rich knowledge and powerful sequence modeling capabilities of PLMs to enhance SR.
Some works propose to use them as text encoders (e.g., build item representations from their textual representations based on PLMs rather than merely using the conventional random initialized ID embeddings). For example, IDA-SR \cite{IDA-SR} and UniSRec \cite{UniRec} apply BERT to obtain item representations from their corresponding texts. LLM2BERT4Rec \cite{LLM-SR} utilizes an embedding model with the titles of the items to retrieve their embeddings. Besides, some researchers try to convert users' historical interactions and items into plain text and utilize them to fine-tune the PLMs for SR \cite{qu2023thoroughly,M6-rec,li2023text,geng2022recommendation}.

Recently, the emergence of Large Language Models (LLMs) \cite{GPT3,gpt4,instruction-tuning} has revolutionized the field of natural language processing with their powerful reasoning abilities and rich world knowledge. Therefore, some pioneer works explore the possibility of directly using powerful LLMs for recommendation via prompt or in-context learning \cite{hou2023large,gao2023chat,wang2023zero,liu2023chatgpt,kang2023llms}. Although these methods achieve impressive results in cold-start scenarios, they still perform poorly in SR when interactions are sufficient since behavioral information is essential for making precise recommendations.  Another line of work considers introducing collaborative signals into LLMs to improve their recommendation performance. TallRec \cite{bao2023tallrec} and InstructRec \cite{zhang2023recommendation} perform instruct-tuning on LLMs to adapt them for understanding behavioral patterns. RLMRec \cite{ren2023representation} aligns the semantic space of LLMs with
collaborative relational signals through cross-view alignment to improve the quality of representations. BIGRec \cite{bao2023bi} treats SR as a text generation task, generating item titles by a two-step approach with the LLM fine-tuned on user behavioral sequences. Nevertheless, these methods lack a thorough examination of whether PLMs' capabilities are being effectively and efficiently harnessed. In our paper, we deeply explore the involvement of PLMs in SR tasks. We intend to scrutinize whether their formidable abilities are truly being utilized to their full potential and offer insights on how to make the most out of PLMs in the context of SR.

\section{Analyses on Behavior Sequence Modeling of PLM-based SR Models}
\label{analysis}

In this section, we comprehensively analyze the capability and suitability of PLM-based SR models on behavior sequence modeling. Without loss of generality, we employ the SOTA PLM-based SR model RECFORMER \cite{li2023text} for the following explorations.

\subsection{Background of RECFORMER}

RECFORMER is a PLM-based SR model that comprises a classical PLM Longformer \cite{longformer} and two additional embeddings that indicate the type of token (determining the token belongs to the special token [CLS], textual attribute key or textual attribute value) and its associated item (which item the token belongs to). It transforms items and behavioral sequences into text sequences and then utilizes the behavior-tuned model to obtain their corresponding representations for providing recommendations. RECFORMER's training process has three steps, including pre-training and two-stage fine-tuning. To simplify, we designate these stages as \emph{stage-PT}, \emph{stage-FT1}, and \emph{stage-FT2} respectively. 
(1) In stage-PT, RECFORMER adopts Masked Language Modeling (MLM) and item-item contrastive (IIC) tasks to pre-train the model. Considering the computational costs, RECFORMER treats in-batch next items as negative instances instead of negative sampling or fully softmax for the IIC task.
(2) In stage-FT1, RECFORMER encodes all items before each epoch starts, obtaining their corresponding item embeddings, then fine-tuning all parameters with these frozen embeddings (for efficiency considering the large size of item candidates).
(3) In stage-FT2, the RECFORMER selects the optimal model from stage-FT1 to acquire its item embeddings and fix them. Stage-FT2 only tunes the parameters of the behavior sequence modeling part (similar to stage-FT1, but with the item embeddings remaining fixed and not updated each epoch).
This strategy of maintaining fixed item embeddings is primarily motivated by the significant computational demands associated with generating all item representations during the cross-entropy loss calculation. 
Therefore, the fine-tuning process of RECFORMER could be simplified as: (1) obtaining the item embeddings from behavior-tuned Longformer, and (2) tuning PLM solely for behavior sequence modeling with fixed item embeddings. The latter step is our focus.

\subsection{Analysis on Attention Discrepancy}

\begin{figure*}[!hbpt]
\centering  
\subfigure[\textbf{Longformer-Instruments}]{
\includegraphics[width=8.8cm,height = 5.2cm]{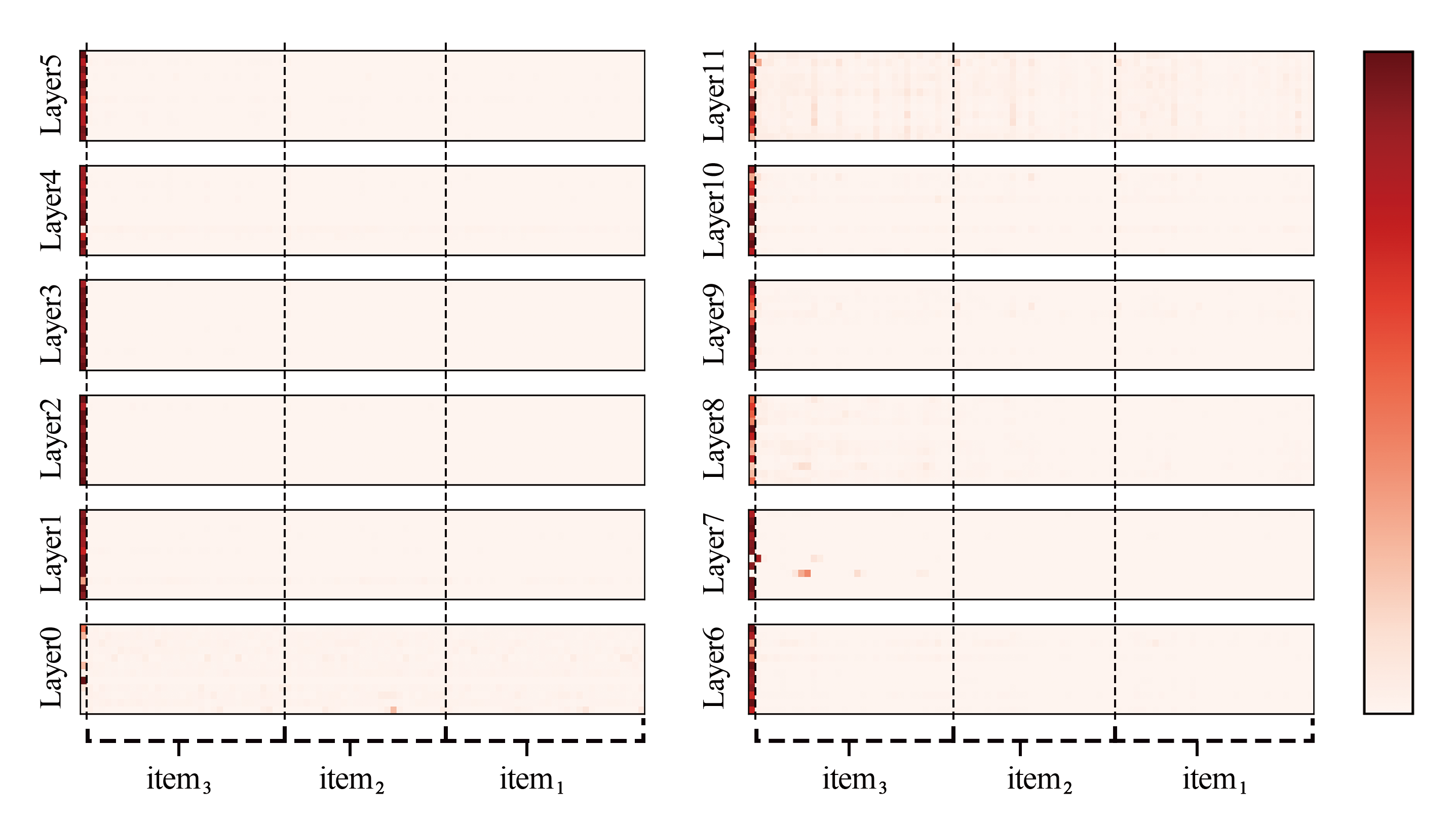}}
\hspace{-0.5cm}
\subfigure[\textbf{RECFORMER-Instruments}]{
\includegraphics[width=8.8cm,height = 5.2cm]{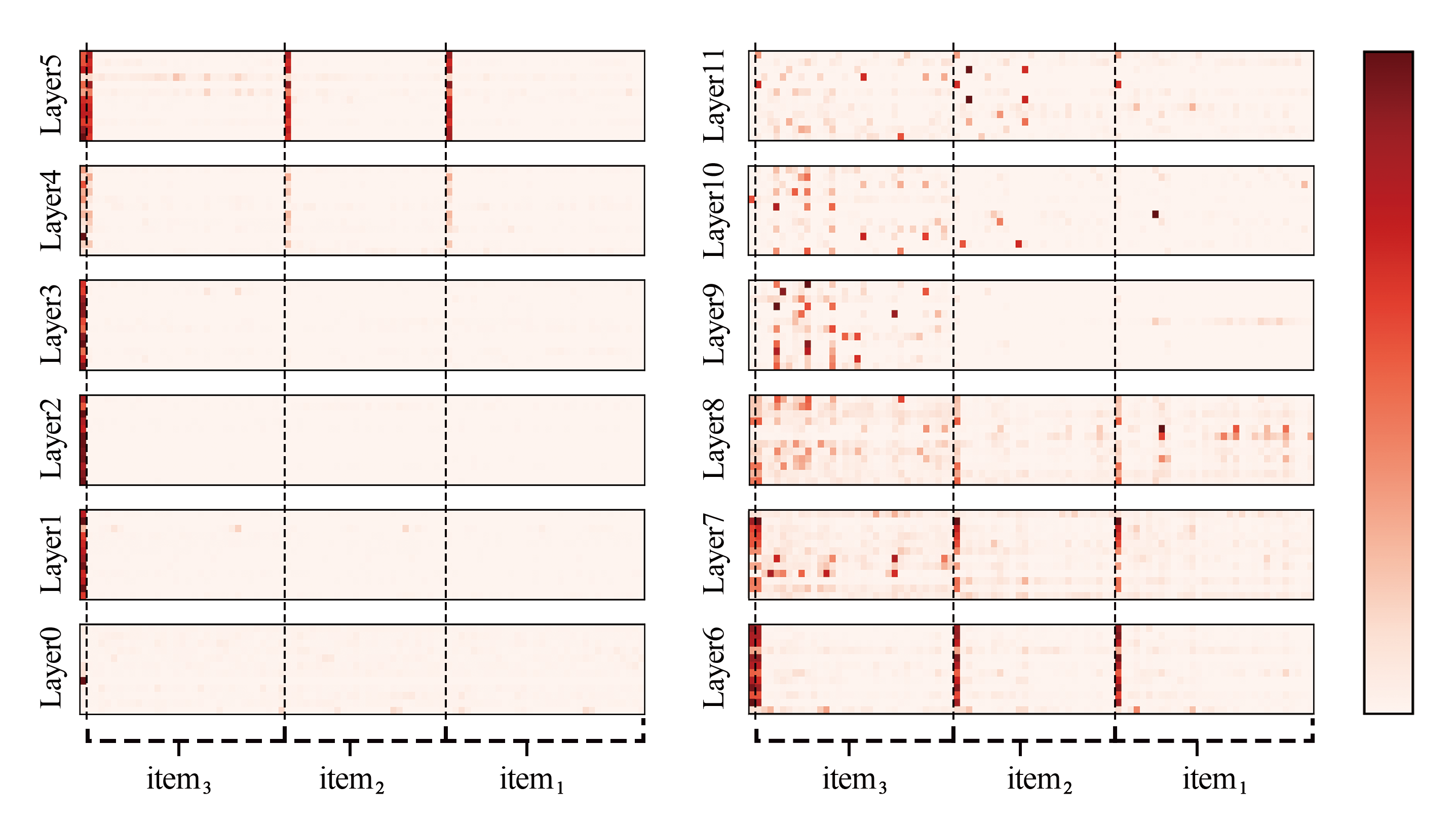}}

\subfigure[\textbf{Longformer-Pantry}]{
\includegraphics[width=8.8cm,height = 5.2cm]{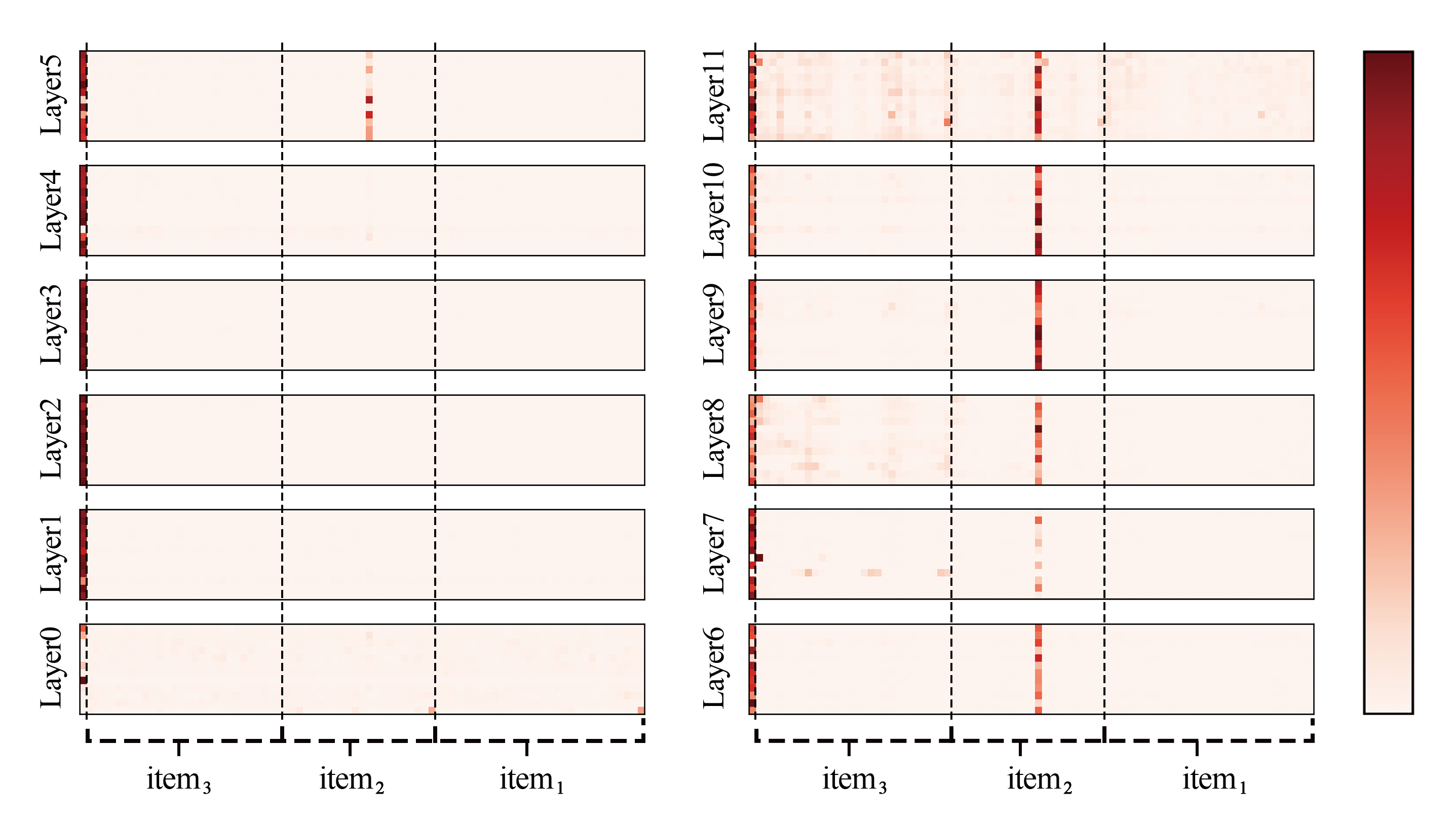}}
\hspace{-0.5cm}
\subfigure[\textbf{RECFORMER-Pantry}]{
\includegraphics[width=8.8cm,height = 5.2cm]{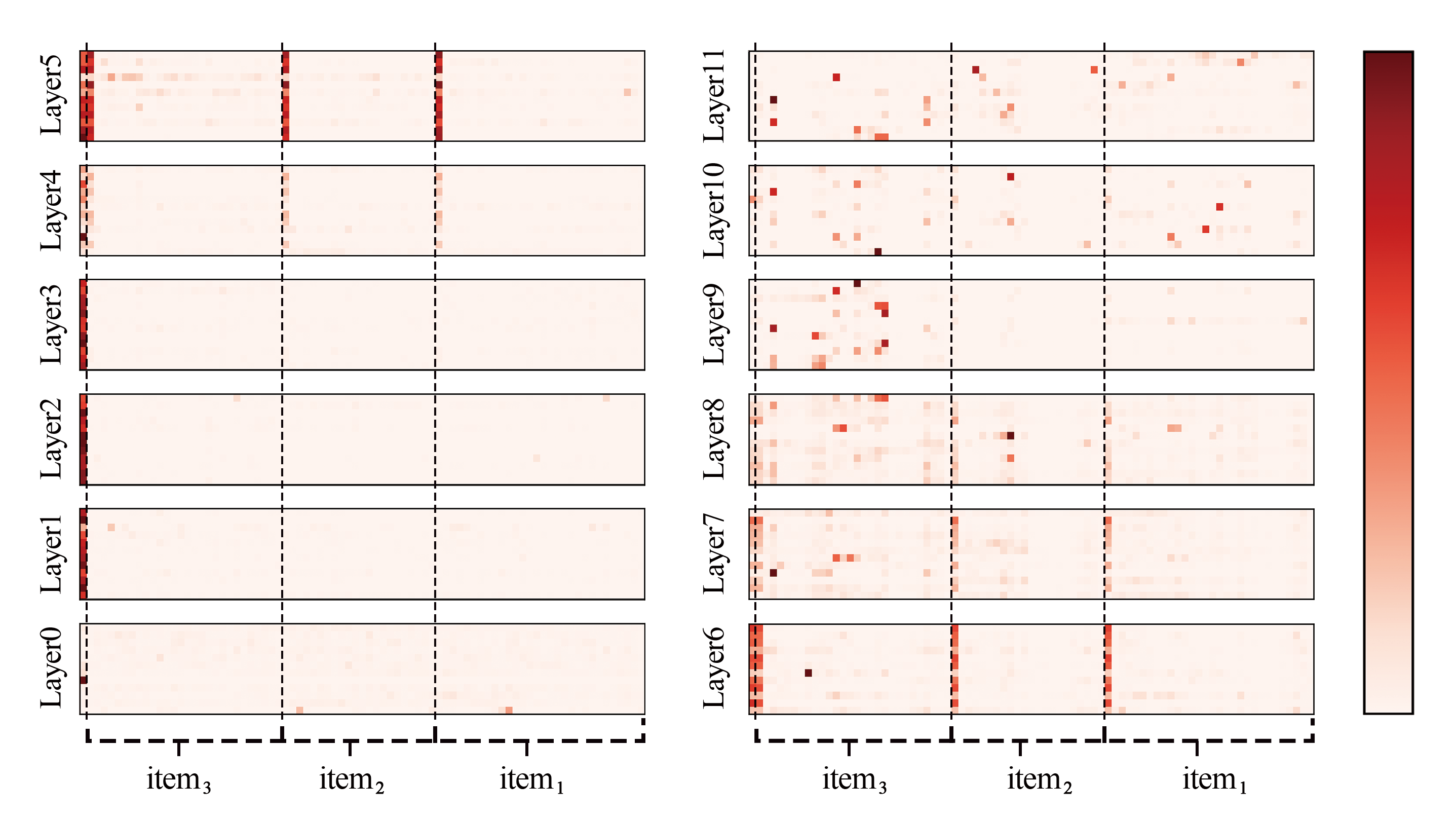}}

\caption{Attention maps of [CLS] in Longformer and RECFORMER. The horizontal axis represents the tokens of items, and the vertical axis represents the attention heads. They have different sequence modeling patterns.}
\label{fig:attention}
\end{figure*}

Intuitively, the sequential patterns of items in SR and tokens in LM significantly differ. SR tends to prioritize the most recent interactions of users, focusing more on their immediate past behavior to make predictions \cite{SASRec, BERT4Rec}. On the other hand, LM concentrates more on maintaining linguistic consistency and adhering to the syntactic structure of the text, ensuring coherence and grammatical accuracy throughout the content. To verify this assumption, we visualize the attention distributions of the special token [CLS] that is used to get the user representation for recommendation in RECFORMER (and its backbone Longformer \cite{longformer} for comparisons).

Specifically, we follow the setting in \cite{li2023text} to pre-train and fine-tune RECFORMER, and stack the attention scores from different heads in the same layer on the vertical axis. As for Longformer, we directly utilize its original checkpoint, drawing the attention maps in the same way as RECFORMER. It's important to highlight that RECFORMER reverses the user's interaction sequence. Consequently, within the attention maps, items positioned to the left symbolize those with which the user has most recently interacted. In Fig. \ref{fig:attention}, we provide the attention maps of two instances from Pantry and Instruments \cite{amazon} with its length of historical sequence as three (for better visualization). Upon examining various instances from multiple domains, we have some universal findings: 

(1) There is a \emph{\textbf{distinct stratification}} in RECFORMER's attention distributions (i.e., the attention distributions are similar within layers 0-3/4-7/8-11) across different instances and domains. 

(2) The attention of RECFORMER's Layers 4-7 significantly focuses on the \emph{\textbf{first token of each item}}, revealing its function as modeling the boundaries between different items. Besides, we also find a universal phenomenon that these first tokens of each item mainly \emph{\textbf{focus on the tokens inside each item at shallow layers}}.
These are obvious SR patterns that resemble the process of generating the corresponding item embeddings of each item in the behavior sequence, which cannot be found in Longformer.

(3) At Layers 8-11, RECFORMER's [CLS] begins to find fine-grained highlight tokens in each item. We find that the highlights of RECFORMER on items are \emph{\textbf{very similar to those of SASRec}} (i.e., more focused on recent items). Therefore, Layers 8-11 might take a role similar to the sequential behavior modeling in classical ID-based SR models.

(4) There is a notable presence of almost the same attention distributions across certain layers and attention heads, indicating \emph{\textbf{significant parameter redundancy}} within PLM-based SR models.

These phenomena are widely-existed in different instances and datasets. In conclusion, we observe that RECFORMER's sequential modeling pattern is totally different from its backbone Longformer's, but similar to conventional ID-based SR models (e.g., SASRec) to some extent. And significant degree of similarity in attention patterns is observed across various attention heads and adjacent layers within the same functional stratification. These observations imply the \textbf{underutilization and parameter redundancy} of the current usage of PLMs in behavior sequence modeling.

\subsection{Analysis on Model Redundancy}

Conventional ID-based SR models typically feature a simplified architecture with fewer layers and heads (for example, SASRec employs 2 layers and 1 head). As reported in their respective studies \cite{SASRec,BERT4Rec}, augmenting the number of attention heads and layers tends to result in a marginal performance improvement and, in some cases, can even cause a reduction in effectiveness. However, PLMs usually have several times more layers and heads (e.g., Longformer has 12 layers and 12 heads), aiming to capture the complex grammar and semantic knowledge within natural language. This level of complexity might be over-qualified for the current requirements of behavior sequence modeling in SR. The observed attention homogenization on different layers and heads also indicates that the current sequential modeling of PLM-based SR models is redundant and could be compressed. Hence, we conduct an intuitive experiment that only tunes several layers of RECFORMER on \emph{Amazon Pantry} as shown in Table \ref{freeze}. We have some findings:

(1) None (i.e., the original Longformer) performs worse than other settings, indicating that behavior-based tuning is essential for obtaining a satisfactory performance. There exists a giant gap between the sequence modeling of PLM and SR.

(2) When the number of tuned layers is fixed, tuning deeper layers always achieves better results in different settings, which may be due to their better task adaption to SR.

(3) By \emph{\textbf{selectively tuning one layer of each stratification}} (e.g., tuning layers 3,7,11), we achieve astonishing results that are comparable to or even exceed those of full fine-tuning in the original RECFORMER. Additionally, we also evaluate on \emph{Arts} and \emph{Instruments}, where merely tuning Layers 3,7,11 could match or closely approach the performance of full fine-tuning. Specifically, we observed changes of HR@5 -0.07\% and NDCG@5 +0.90\% on Instruments, and HR@5 -0.62\%, NDCG@5 -0.18\% on Arts.
It not only confirms the functional segmentation but also highlights the existence of substantial parameter redundancy in PLM for SR. These results imply that the power of PLM in modeling token sequences is not fully activated in existing PLM-based SR models (or is not perfectly suitable for the behavior sequence modeling in SR). 

Based on the above findings, we begin to rethink the current usage of PLMs in SR, ambitiously aiming to investigate the following two problems:
(1) \emph{Can we use more simplified sequential models (e.g., classical ID-based SR models such as SASRec and BERT4Rec) to replace the cumbersome PLMs for more economical behavior sequence modeling?}
(2) \emph{How to effectively and efficiently activate the magic power of PLMs for SR without much inference cost}?

\begin{table}[!hbtp]
\small
\caption{Results of different tuned layers in RECFORMER.}
\label{freeze}
\setlength\tabcolsep{3.3pt}
\begin{tabular}{c |c c |c c c c |c c c}
\toprule
                 \multirow{2}{*}{Metric}  & \multicolumn{9}{c}{Tuned Layers} \\
\cline{2-10}
 & None &All  &0,4,8 & 1,5,9 & 2,6,10 & 3,7,11 & 3 & 7 & 11\\

\midrule     
                          
                           H@5 &23.61	&31.37	&29.90	&31.53	&31.20	&\textbf{32.55}	&28.97	&29.76	&30.76      \\ 
                           H@10 &30.97	&41.08	&39.41	&41.37	&40.74	&\textbf{42.32}	&38.42	&39.68	&40.50\\
                            N@5  &18.56	&24.41	&23.27	&24.32	&24.26	&\textbf{25.27}	&22.60	&23.06	&23.94      \\ 
                            N@10 &20.92	&27.54	&26.33	&27.49	&27.33	&\textbf{28.42}	&25.64	&26.26	&27.07\\
                      
\bottomrule
\end{tabular}
\end{table}

\section{EXPERIMENTS}

In this section, we aim to answer the following research questions:
(RQ1): Does the powerful sequence modeling capability of PLMs effectively benefit SR for better modeling the behavior sequence? If not, could we replace it with a simpler sequence model?
(RQ2): What is the optimal approach to leverage the formidable capabilities of PLMs in SR, especially considering their limitations in modeling sequences? 
(RQ3): What does an ideal framework for integrating PLMs into SR look like, one that harnesses the powerful capabilities of PLMs while maintaining relatively low training and inference costs?
To investigate the above problems, we implement several simplified but effective PLM-based SR variants for exploration, regarding \emph{behavior-tuned PLMs as item initializer} with \emph{simplified classical ID-based behavior sequence modeling}.

\subsection{Behavior-tuned PLMs as Item Initializer}

\label{varient}

We conduct extensive experiments on the SOTA PLM-based SR model RECFORMER \cite{li2023text}, and implement our enhanced variants with simplified sequence modeling based on it. We adopt two classical SR models SASRec \cite{SASRec} and BERT4Rec \cite{BERT4Rec} as the backbone sequence modeling method. Our PLM-based SR variants can be divided into three groups, namely embedding-init (freeze), embedding-init (trainable), and further-train.
Note that we use the behavior-tuned PLM (i.e., Longformer tuned in RECFORMER's certain stages in Sec. 3.1) to initialize item embeddings in our methods.

(a) For \textbf{embedding-init (freeze)}, we experiment with three versions of RECFORMER: the original Longformer \cite{longformer} (LF), RECFORMER trained on all pre-training datasets (PT), and RECFORMER after stage-FT1 on the target domain (FT). We only use the item representations obtained by these models as the fixed item ID embeddings of ID-based SR models (e.g., SASRec), and fine-tune the sequence modeling part of ID-based SR models via the target domain's training data. This setting could be viewed as an SR model with fixed item embeddings given by Longformer/RECFORMER, aiming to explore whether a simple sequence model from conventional ID-based SR is sufficient for behavior modeling of SR.

(b) For \textbf{embedding-init (trainable)}, we allow the initialized item representations above to be trainable along with the sequence modeling part (i.e., SASRec/BERT4Rec with trainable item embeddings that Longformer/RECFORMER has initialized). It provides insights into the impact of concurrent learning on well-initialized ID embeddings via PLM and simplified sequence modeling.

(c) For \textbf{further-train}, we start from the FT version of embedding-init (freeze) and proceed to optimize (i) all model parameters (All) or (ii) just the item embeddings (Emb). In this setting, our well-initialized item ID embeddings obtained from behavior-tuned PLM are further tuned under a matched and warmed-up sequence modeling component of ID-based SR models.

We should highlight that these $8$ variants are all equipped with (a) \textbf{simplified sequence modeling of ID-based SR models}, and (b) \textbf{item initialization based on behavior-tuned PLM}.
The PLM part is \textbf{NOT} tuned/used in the following tuning and serving.

\subsection{Experimental Settings}

\noindent
\textbf{Datasets.}
We conduct experiments on seven real-world datasets from the Amazon review datasets~ \cite{amazon}. 
We use \emph{Books}, \emph{Movies and TV}, \emph{Sports and Outdoors}, \emph{Clothing Shoes and Jewelry} for pre-training, and choose \emph{Arts, Crafts and Sewing}, \emph{Musical Instruments}, and \emph{Pantry} as the downstream domains for evaluation. In addition, we also select Movies and TV and Sports and Outdoors from the pre-training dataset for evaluation to test the impact of pre-training on the source domain. Following \cite{SASRec,BERT4Rec}, for pre-training datasets, we filter out users and items with less $5$ interactions. But for new domain datasets, we just filter out the users with less than 4 interaction records to ensure adequate instances.
We arrange each user's historical behavior sequence by timestamps and adopt leave-one-out setting for building validation and test sets. For each user historical sequence, the last item is used as the test data, the item before the last one is used as the validation data, and the remaining interaction sequences are used for training.

As our research is centered on assessing the efficacy of PLMs in behavior sequence modeling, we are particularly mindful of the potential influence of pre-existing knowledge within PLMs on zero-shot capabilities in conventional SR models. To address this concern and ensure a more accurate evaluation, we have strategically \textbf{\emph{excluded all validation and test instances that contain items not previously trained in the model}}. This step is crucial as it helps to isolate the effects of PLMs on zero-shot scenarios in our analysis and provides a clearer understanding of their true performance in SR tasks, free from the confounding effects of prior knowledge on cold-start items. The detailed statistics are in Table \ref{statistic}.

\noindent
\textbf{Evaluation Metrics.}
To evaluate the performance of the Sequential Recommendation task, we follow the classical setting \cite{SASRec,BERT4Rec} and select $100$ randomly sampled negative items for evaluation. We adopt two widely-used metrics Hit rate (HR@k) and Normalized Discounted Cumulative Gain (NDCG@k) with $k = 5,10$.

\noindent
\textbf{Implementation Details.}
We implement RECFORMER with its provided source code, pre-training and fine-tuning it with the same hyper-parameters mentioned in the original paper. For simplified sequence models, the batch size is set to 256 and 128 for BERT4Rec and SASRec separately, and we carefully tune the learning rate in \{0.0003, 0.001, 0.003, 0.01\}. Additionally, to directly utilize the representations obtained by the RECFORMER, we set the embedding dimension as 768 for SASRec and BERT4Rec, which is identical to the RECFORMER's. To ensure a fair comparison, we optimize all the models with Adam optimizer. Early stopping is adopted with the patience of 10 epochs to prevent overfitting, and NDCG@10 is set as the indicator. 
\begin{table}[t]
\small
    \caption{Statistics of seven pre-training domain and new domain datasets.}
    \label{statistic}
    \centering
    \begin{tabular}{l|rrrr}
        \toprule
        \textbf{Dataset} & \# \textbf{User} & \# \textbf{Item} & \# \textbf{Interactions} & \textbf{Avg.len}  \\
        \midrule
        \textbf{Pre-training} & 402,979 & 930,518 & 3,547,017 &8.8  \\
        - Books & 197,891 & 504,085 & 1,990,164 & 10.1  \\
        - Clothing & 135,041 & 294,788 & 1,004,679 & 7.4  \\
        - Sports & 39,477 & 87,235 & 262,998 & 6.7  \\
        - Movies & 30,570 & 44,410 & 289,176 & 9.5  \\
        \midrule
        \textbf{Arts} & 131,149 & 138,116 & 718,628 & 5.5   \\
        \textbf{Instruments} & 62,691 & 53,899 & 403,135 & 6.4  \\
        \textbf{Pantry} & 22,601 & 8,249 & 179,735 & 8.0  \\
        \bottomrule
    \end{tabular}
\end{table}

\begin{table*}[!hbpt] 
\caption{Overall performance comparisons. SAS, BERT, and REC represent the original  SASRec, BERT4Rec, and RECFORMER. LF, PT, FT represent using Longformer, pre-trained RECFORMER, and fine-tuned RECFORMER for embedding initialization. Improv. gives the relative improvements compared to the original base SR models (SASRec/BERT4Rec), which are significant. }
\label{performance}
\begin{tabular}{c|l|p{0.9cm}<{\centering}|p{0.9cm}<{\centering}|p{0.85cm}<{\centering}p{0.85cm}<{\centering}p{0.85cm}<{\centering}|p{0.85cm}<{\centering}p{0.85cm}<{\centering}p{0.85cm}<{\centering}|cc|p{1.2cm}<{\centering}}
\toprule
   \multirow{2}{*}{Dataset} & \multirow{2}{*}{Metric}  &  \multirow{2}{*}{SAS} &  \multirow{2}{*}{REC}& \multicolumn{3}{c|}{Embedding-init (freeze) }  &\multicolumn{3}{c|}{Embedding-init (trainable)}& \multicolumn{2}{c|}{Further-train} &\multirow{2}{*}{Improv.}\\
\cline{5-12}

   & & &  & LF & PT  & FT & LF & PT  & FT &All &Emb &  \\
\midrule

 \multirow{4}{*}{Arts}

                             & H@5        &58.51	&59.60	&44.23	&56.61	&59.66	&58.13	&67.75	&70.85	&\textbf{71.29}	&71.01 &\textbf{+21.84\%}\\
                            & H@10        &68.54	&68.72	&56.57	&67.76	&70.40	&69.00	&76.78  &80.70	&80.62	&\textbf{81.05} &\textbf{+18.25\%}\\
                             & N@5      &48.35	&49.88	&33.33	&45.16	&48.63	&47.46	&55.27	&58.86	&\textbf{59.72}	&59.13 &\textbf{+23.52\%} \\
                             &N@10 &51.59	&52.48	&37.32	&48.74	&52.10	&50.97	&58.31	&62.01	&\textbf{62.78}	&62.27 &\textbf{+21.69\%}\\
                            
                            \midrule

 \multirow{4}{*}{Instruments}

                             & H@5        &58.42	&53.18	&43.84	&54.22	&54.49	&58.73	&66.06	&67.28	&68.28	&\textbf{68.37}  &\textbf{+17.03\%}      \\
                             & H@10        &68.10	&63.70	&56.19	&66.37	&65.97	&69.05	&75.41	&77.61	&\textbf{78.55}	&78.06 &\textbf{+15.35\%}  \\
                             & N@5      &48.66	&44.06	&33.16	&43.10	&43.63	&48.64	&55.27	&55.14	&\textbf{56.35}	&56.34   &\textbf{+15.80\%} \\
                             & N@10       &51.74	&47.46	&37.15	&47.02	&47.48	&52.02	&58.31	&58.50	&\textbf{59.68}	&59.51 &\textbf{+15.35\%} \\
                            \midrule

 \multirow{4}{*}{Pantry}

                             & H@5        &31.80	&31.37	&20.50	&31.51	&30.42	&31.78	&\textbf{37.85}	&33.03	&36.39	&35.92   &\textbf{+19.03\%}     \\
                             & H@10 &43.48	&41.08	&30.93	&42.43	&40.91	&44.12	&\textbf{49.81}	&45.25	&48.28	&48.20  &\textbf{+14.56\%}\\
                             & N@5      &23.62	&24.41	&13.73	&23.78	&23.00	&22.47	&\textbf{28.71}	&23.90	&27.48	&27.02   &\textbf{+21.55\%}  \\
                             & N@10 &27.40	&27.54	&17.08	&27.29	&26.35	&26.34	&\textbf{32.56}	&27.83	&31.31	&31.01 &\textbf{+18.83\%} \\
                            \midrule

 \multirow{4}{*}{Sports}

                             & H@5        &41.76	&45.89	&37.73	&47.03	&46.28	&39.90	&48.88	&54.64	&54.68	&\textbf{56.08} &\textbf{+34.29\%}       \\
                            & H@10 &52.16	&57.26	&49.01	&59.42	&58.21	&51.29	&59.69	&66.36	&66.53	&\textbf{67.84} &\textbf{+30.06\%}\\
                             & N@5      &33.42	&36.86	&28.55	&36.59	&36.31	&31.42	&39.05	&43.55	&43.51	&\textbf{44.70} &\textbf{+33.75\%}    \\
                             & N@10 &36.84	&40.53	&32.17	&40.59	&40.03	&35.09	&42.54	&47.33	&47.27	&\textbf{48.50} & \textbf{+31.65\%} \\

                            \midrule

 \multirow{4}{*}{Movies}

                             & H@5  &61.58	&53.90	&39.87	&56.30	&55.46	&61.14	&68.86	&68.36	&68.19	&\textbf{69.48}&\textbf{+12.83\%}           \\
                            & H@10 &70.52	&63.67	&52.04	&67.26	&66.12	&71.11	&77.96	&78.07	&77.30	&\textbf{79.02} &\textbf{+12.05\%}          \\
                             & N@5 &51.42	&45.27	&28.78	&45.40	&45.29	&50.55	&57.50	&56.38	&56.86	&\textbf{57.71}   &\textbf{+12.23\%}      \\
                            & N@5 &54.30	&48.42	&32.71	&48.95	&48.73	&53.77	&60.45	&59.53	&59.81	&\textbf{60.81}  &\textbf{+11.99\%} \\

\bottomrule
\end{tabular}

\begin{tabular}{c|l|p{0.9cm}<{\centering}|p{0.9cm}<{\centering}|p{0.85cm}<{\centering}p{0.85cm}<{\centering}p{0.85cm}<{\centering}|p{0.85cm}<{\centering}p{0.85cm}<{\centering}p{0.85cm}<{\centering}|cc|p{1.2cm}<{\centering}}
\toprule
   \multirow{2}{*}{Dataset} & \multirow{2}{*}{Metric}  &  \multirow{2}{*}{BERT} &  \multirow{2}{*}{REC}& \multicolumn{3}{c|}{Embedding-init (freeze) }  &\multicolumn{3}{c|}{Embedding-init (trainable)}& \multicolumn{2}{c|}{Further-train} &\multirow{2}{*}{Improv.}\\
\cline{5-12}

   & & &  & LF & PT  & FT & LF & PT  & FT &All &Emb &  \\
\midrule

 \multirow{4}{*}{Arts}

                             & H@5 &52.41	&59.60	&48.76	&61.81	&64.17	&54.91	&62.38	&64.26	&62.35	&\textbf{64.64}  &\textbf{+23.34\%}      \\
                            & H@10 &62.34	&68.72	&60.66	&71.33	&73.24	&65.75	&71.99	&73.29	&70.95	&\textbf{73.50} &\textbf{+17.90\%} \\
                             & N@5   &43.22	&49.88	&37.36	&51.24	&54.12	&44.30	&51.86	&54.23	&52.85	&\textbf{54.66} &\textbf{+26.47\%}   \\
                             & N@10 &46.43	&52.48	&41.21	&54.33	&57.07	&47.81	&55.00	&57.16	&55.64  &\textbf{57.53}   &\textbf{+23.91\%}  \\
                            \midrule

 \multirow{4}{*}{Instruments}

                             & H@5    &54.20	&53.18	&49.16	&61.17	&63.61	&54.95	&61.75	&63.51	&62.23	&\textbf{64.23}    &\textbf{+18.51\%}        \\
                            & H@10 &63.51	&63.70	&59.12	&71.04	&72.86	&65.74	&71.50	&72.88	&71.07	&\textbf{73.44} &\textbf{+15.64\%} \\
                             & N@5  &45.53	&44.06	&39.52	&50.83	&53.65	&44.70	&51.69	&53.68	&52.64	&\textbf{54.38}    &\textbf{+19.44\%}    \\
                            & N@10 &48.54	&47.46	&42.74	&54.02	&56.66	&48.09	&54.85	&56.71	&55.50	&\textbf{57.36} &\textbf{+18.17\%}\\
                            \midrule

 \multirow{4}{*}{Pantry}

                             & H@5      &28.93	&31.37	&24.49	&30.71	&33.57	&27.23	&32.77	&34.43	&33.50	&\textbf{34.75}    &\textbf{+20.12\%}     \\
                            &H@10 &39.91	&41.08	&36.48	&42.36	&45.01	&39.61	&44.62	&45.59	&44.66	&\textbf{46.11} &\textbf{+15.53\%}   \\
                             & N@5    &21.51	&24.41	&16.76	&22.58	&25.26	&19.55	&24.24	&25.87	&25.09	&\textbf{26.24} &\textbf{+21.99\%}      \\
                             &N@10 &25.03	&27.54	&20.57	&26.32	&28.94	&23.52	&28.05	&29.48	&28.69	&\textbf{29.89} & \textbf{+19.42\%}     \\

                            \midrule

 \multirow{4}{*}{Sports}

                             & H@5   &31.12	&\textbf{45.89}	&30.87	&43.68	&43.23	&33.19	&43.83	&43.11	&42.98	&43.47    &\textbf{+40.84\%}        \\
                            & H@10 &41.28	&\textbf{57.26}	&41.08	&54.91	&53.95	&43.55	&55.02	&54.10	&53.28	&54.10 &\textbf{+33.28\%}\\
                             & N@5   &24.51	&\textbf{36.86}	&22.75	&34.29	&34.05	&25.24	&34.59	&34.15	&34.10	&34.51   &\textbf{+41.13\%}     \\
                            &N@10 &27.78	&\textbf{40.53}	&26.04	&37.91	&37.51	&28.58	&38.20	&37.69	&37.42	&37.94 &\textbf{+37.51\%}\\
                            \midrule

 \multirow{4}{*}{Movies}

                             & H@5   &56.67	&53.90	&46.00	&62.68	&64.71	&56.44	&63.74	&65.12	&63.04	&\textbf{65.83}    &\textbf{+16.16\%}     \\
                            & H@10 &66.08	&63.67	&56.79	&72.58	&74.26	&67.33	&73.48	&74.53	&71.97	&\textbf{74.99} &\textbf{+13.48\%}\\
                             & N@5  &47.32	&45.27	&33.77	&51.76	&53.06	&44.65	&53.12	&54.46	&52.71	&\textbf{55.33}  &\textbf{+16.93\%}      \\
                            &N@10 &50.35	&48.42	&37.77	&54.97	&57.19	&48.18	&56.23	&57.50	&55.60	&\textbf{58.29} &\textbf{+15.77\%}\\

\bottomrule
\end{tabular}
\label{tab:main}
\end{table*}

\subsection{Main Results}
\label{main}

Table \ref{performance} shows the overall results on five datasets and two simplified SR models, we have the following astonishing findings:

(1) \textbf{\emph{PLM's powerful sequence modeling capability is not fully activated or overqualified when performing behavior modeling in SR. The simple sequence modeling of ID-based SR performs well enough with better efficiency.}} In general, the FT embedding-init (freeze) has comparable or even better performance across most datasets compared to RECFORMER. It indicates that a simple sequence model is capable of behavior modeling in SR, while the sequence modeling ability of PLMs may be redundant.

(2) \textbf{\emph{Item initialization based on behavior-tuned PLM could substantially boost the performance, while vanilla PLMs' initializations have no effort.}} Models with LF-initialized embeddings exhibit inferior performance compared to the original random initialization. However, both PT and FT initializations consistently lead to significant improvements. This holds true regardless of whether RECFORMER's original performance surpasses that of SASRec and BERT4Rec. It implies that the powerful semantic knowledge in vanilla PLMs has a huge gap with the personalized user preference in SR. Directly integrating PLMs into SR might inadvertently introduce excessive noise, potentially undermining the performance of the model. On the contrary, behavior-tuned PLMs could manage and utilize the knowledge within it effectively when facilitating SR.

(3) \textbf{\emph{Further tuning on item representations given by behavior-tuned PLM could further improve the results.}} Further-train variants achieve the overall best performance, and embedding-init (trainable) outperforms embedding-init (freeze). These two settings conduct further tuning on item representations, which indicates that the item initialization of behavior-tuned PLM requires further adaptation to obtain sufficient behavior information. Moreover, we notice that solely further training item embeddings (i.e., with fixed sequence modeling part) in further-train has the best performance. It indicates that further training embedding might be a more effective and stable way to improve model performance further.

(4) \textbf{\emph{The power of behavior-tuned PLMs as item initializers is potentially transferable.}} Comparing LF, PT, and FT settings, we observe that even PLMs tuned on pre-training datasets (PT) could also bring in essential user preference knowledge and largely benefit SR in the downstream new domains compared to the original ID-based SR models and vanilla PLM initialization. Future research could focus on the novel usage of item initialization from multi-domain behavior-tuned PLMs for various new domains. Furthermore, developing a more efficient framework for pre-training behavior-tuned PLMs is also worthy for researchers to investigate.

(5) \textbf{\emph{Sequence modeling methods that possess analogous architectures and training objectives to those of behavior-tuned PLMs benefit more from our PLM-based initialization.}} Our item initialization is universal for different ID-based sequence modeling. Comparing the results based on SASRec and BERT4Rec, we find that our initialization facilitates more for BERT4Rec under the freeze setting. Besides, its FT (freezing) performs comparably to FT (trainable). We hypothesize that it is attributed to the similarities in sequence modeling architectures and training losses between RECFORMER and BERT4Rec. It also supports our hypothesis that PLMs may be excessively complex for behavior sequence modeling and could be substituted with simpler sequence models.

\subsection{Results of Full-ranking Setting}

The findings in Sec. \ref{main} demonstrate the effectiveness of our proposed behavior-tuned PLMs initialized item embeddings under the random negative sampling setting. To further assess the robustness of our method, we conduct experiments with the full-ranking setting \cite{metric,metric2} (i.e., ranking the positive item with the whole item set) on the same datasets.
From Table \ref{full-ranking}, we have:

(1) Adopting our behavior-tuned PLMs initialized item embeddings could still greatly improve the performance of original SASRec under the full-ranking setting, while more improvements are made in metrics related to coarse-grained accuracy (HR@k). It indicates that our proposed method has more advantages in improving the model's generalization ability (i.e., tends to push positive samples to a relatively higher rank among all candidates, rather than the specific top 1 position).

(2) Our method still consistently outperforms RECFORMER on all datasets. Compared with the random sampling setting in Table \ref{tab:main}, the improvement in the full-ranking setting is not that impressive. This phenomenon is consistent with previous works \cite{li2023text,qu2023thoroughly} that PLMs-based SR models perform well at making precise recommendations (i.e., pushing positive samples to the very top part of all candidates). It should be highlighted that our method greatly reduces the training and inference costs compared with RECFORMER. We only use PLM for item initialization, and the subsequent tuning and inference are only based on simplified sequence models without PLM.
Considering our simple structure in serving (the same as SASRec), the current improvement is acceptable. The advantage of our method might be more activated in the matching stage (selecting the top hundreds of items from millions of candidates).

\begin{table}[!hbtp]
\small
\caption{Results of RECFORMER, original SASRec, and FT Embedding-init (trainable) SASRec under the full-ranking setting (\%). Improv. (REC) and Improv. (SAS) represent the relative improvements compared with the RECFORMER and original SASRec respectively.}
\label{full-ranking}
\setlength\tabcolsep{3.3pt}
\begin{tabular}{c |c |c |c | c|c c}
\toprule
                 \multirow{2}{*}{Dataset} & \multirow{2}{*}{Metric} & \multirow{2}{*}{REC} & \multirow{2}{*}{SAS}  & \multirow{2}{*}{FT-init } & \multicolumn{2}{c}{Improv.}   \\
                 \cline{6-7}
                  & & &   &   & REC & SAS\\
    
  \midrule
                 \multirow{6}{*}{Arts}  &H@5  &13.83  &13.62 & 14.46 &+4.56\% &+6.17\%   \\     
                                        &H@10 &15.61	&15.21	&16.94 &+8.52\% &+11.37\%\\
                                        &H@50 &21.02	&20.46	&24.69 &+17.46\% &+20.67\% \\
                                        &N@5  &12.09  &11.84	 &12.29 &+1.65\% &+3.80\%\\
                                        &N@10 &12.66	&12.35	&13.09 &+3.40\% &+6.00\% \\
                                        &N@50 &13.86	&13.50	&14.73  &+6.28\% & +9.11\%\\
\midrule
 \multirow{6}{*}{Instruments} &H@5  &20.00	&20.04	&20.71 & +3.55\% & +3.34\%  \\       
        &H@10 &20.99	&21.75	&22.68 &+8.05\% &+4.28\%\\
        &H@50 &24.49	&27.71	&29.50 &+20.46\% &+6.46\% \\
      &N@5  &18.83	&18.28	&18.92 & +0.48\% & +3.50\%\\
      &N@10 &19.15	&18.83	&19.54 &+2.04\% &+3.77\% \\
      &N@50 &20.18	&20.13	&20.92 &+3.67\% &+3.92\%\\
      \midrule
 \multirow{6}{*}{Pantry} &H@5  &7.75 &7.04 &9.10 & +17.42\% & +29.26\%  \\       
        &H@10 &9.97	&8.37	&11.45 &+14.84\% &+36.80\%\\
        &H@50 &15.76	&13.25	&18.31 &+16.18\% &+38.19\% \\
      &N@5  &6.13 &5.87 &6.69 & +9.14\% & +13.97\%\\
      &N@10 &6.85	&6.30	&7.45 &+8.76\% &+18.25\% \\
      &N@50 &8.12	&7.35	&8.96 &+10.34\% &+21.90\%\\
      \midrule
 \multirow{6}{*}{Sports} &H@5  &12.49 &11.00 &12.99 &+4.00\% & +18.09\%     \\    
        &H@10 &13.47	&11.65	&14.07 &+4.45\% &+20.77\% \\
        &H@50 &16.19	&13.27	&17.36 &+7.23\% &+30.82\%\\
      &N@5  &11.10 &9.88 &11.12 &+0.18\% & +12.55\%\\
      &N@10 &11.43	&10.09	&11.46 &+0.26\% &+13.58\%\\
      &N@50 &12.09	&10.44	&12.18 &+0.74\% &+16.67\%\\
 \midrule
 \multirow{6}{*}{Movies} 
 &H@5  &21.48	&21.43	&21.97 &+2.28\% & +2.52\%     \\    
        &H@10 &23.10	&23.48	&23.84 &+3.20\% &+1.53\% \\
        &H@50 &27.50	&30.33	&29.95 &+8.91\% &-\\
      &N@5  &19.36	&19.05	&19.56 &+1.03\% & +2.68\%\\
      &N@10 &19.89	&19.71	&20.15 &+1.31\% &+2.23\%\\
      &N@50 &20.85	&21.20	&21.47 &+2.97\% &+1.27\%\\
                      
\bottomrule
\end{tabular}
\end{table}

\subsection{Analysis on Further Cooperations}
\label{further exploraction}
To further explore the effectiveness and robustness of our proposed behavior-tuned PLMs initialized embeddings for SR, we conduct two variants of our item initialization for evaluation.

\noindent
\textbf{Cooperating with trainable ID embeddings.}
Some previous works enhance item representation by adding the textual embeddings to its trainable item ID embeddings \cite{UniRec,IDA-SR}. Following this, we regard our behavior-tuned PLM initialization as the extra textual embeddings in SASRec and fix it during training (i.e., initialized embeddings as additional training signals). The results in Table \ref{further} demonstrate that employing behavior-tuned PLMs initialized embeddings (i.e., PT and FT) leads to substantial performance enhancements, while embeddings from original PLMs (LF) do not. The results reconfirm that it is the behavior-aware semantics, not the original semantic information, that improves the SR performance.

\noindent
\textbf{Cooperating with RECFORMER's sequence modeling.}
In this evaluation, we set the item embeddings of RECFORMER in its stage-FT2 to be trainable, with other settings unchanged.
This cooperation significantly improves over the original RECFORMER on all datasets, indicating that the behavior-tuned PLMs initialized embeddings are beneficial even with PLM-based sequence modeling. However, its performance is still slightly inferior with using SASRec's behavior sequence modeling. It validates the effectiveness of our PLM-based item initialization and further highlights the redundancy of PLMs in behavior sequence modeling. Besides, these results also reveal insufficient behavior information within item representations that are directly obtained from behavior-tuned PLMs. We believe that due to the huge conflicts between the semantic and behavioral representations of items, the superior strategy of representing items with PLM is: \textbf{\emph{do not strongly coupled with texts, but using behavior-tuned PLMs to build item initialization}}, which highlights behavioral information while also adopting relevant semantic information as an important supplement.

\begin{table}[!hbtp]
\small
\caption{Results of different variants for using our proposed behavior-tuned PLMs initialized embeddings (\%).}
\label{further}
\setlength\tabcolsep{3.3pt}
\begin{tabular}{c |c |c |c c c |c c}
\toprule
                 \multirow{2}{*}{Dataset} & \multirow{2}{*}{Metric} & \multirow{2}{*}{SASRec} & \multicolumn{3}{c|}{with trainable ID}  & \multicolumn{2}{c}{RECFORMER} \\
                 
\cline{4-8}
  & & & LF  &PT & FT & original & trainable\\
  \midrule
                 \multirow{4}{*}{Arts}  &H@5  &58.51  &59.04	&67.81	&71.55	&59.60  &68.70      \\    

                                        &H@10 &68.54 &69.79	&77.74	&80.91&68.72	&78.01\\
                                        &N@5  &48.35  &48.30	&55.58	&59.83	&49.88  &58.36\\
            &N@10 &51.59 &51.77	&58.80	&62.87	& 52.48&61.37\\
\midrule

 \multirow{4}{*}{Pantry} &H@5  &31.80  &31.71	&38.47	&38.32	&31.37  &36.71   \\   
 & H@10 &43.48 &43.75	&50.61	&50.38	&41.08&47.82\\
      &N@5  &23.62  &22.46	&29.17	&28.88	&24.41  &28.08\\
&N@10 &27.40 &26.43	&33.10	&32.78	&27.54&31.66\\
\midrule

 \multirow{4}{*}{Sports} &H@5  &41.76  &39.83	&53.95	&50.32	&45.89  &49.84     \\               
 &H@10 &52.16 &51.72	&65.74	&62.10	& 57.26&61.34\\
      &N@5  &33.42  &29.16	&42.58	&38.78	&36.86  &40.34\\
&N@10 &36.84 &32.99	&46.40	&42.59	&40.53&44.06\\

\bottomrule
\end{tabular}
\end{table}

\begin{figure}[!hbpt]
\centering  
\subfigure[\textbf{Instruments}]{
\includegraphics[width=8.4cm,height = 3.3cm]{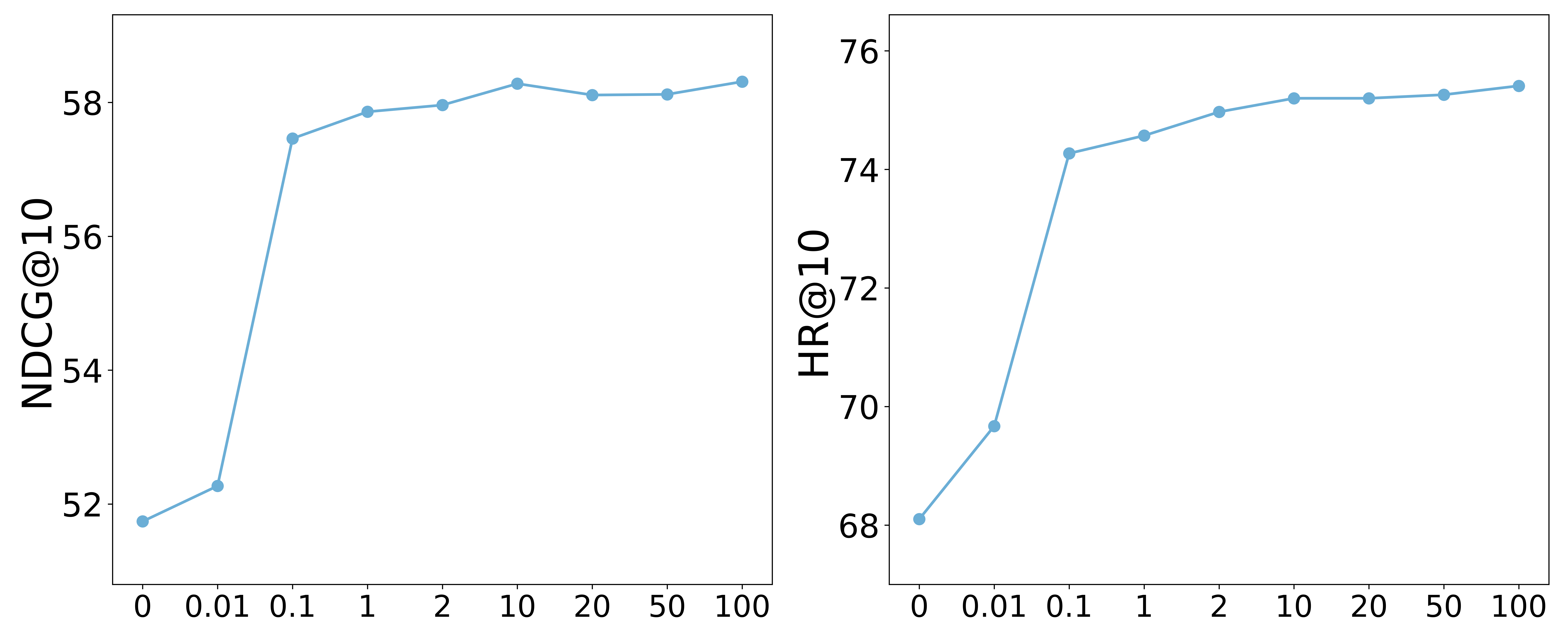}}

\subfigure[\textbf{Pantry}]{
\includegraphics[width=8.4cm,height = 3.3cm]{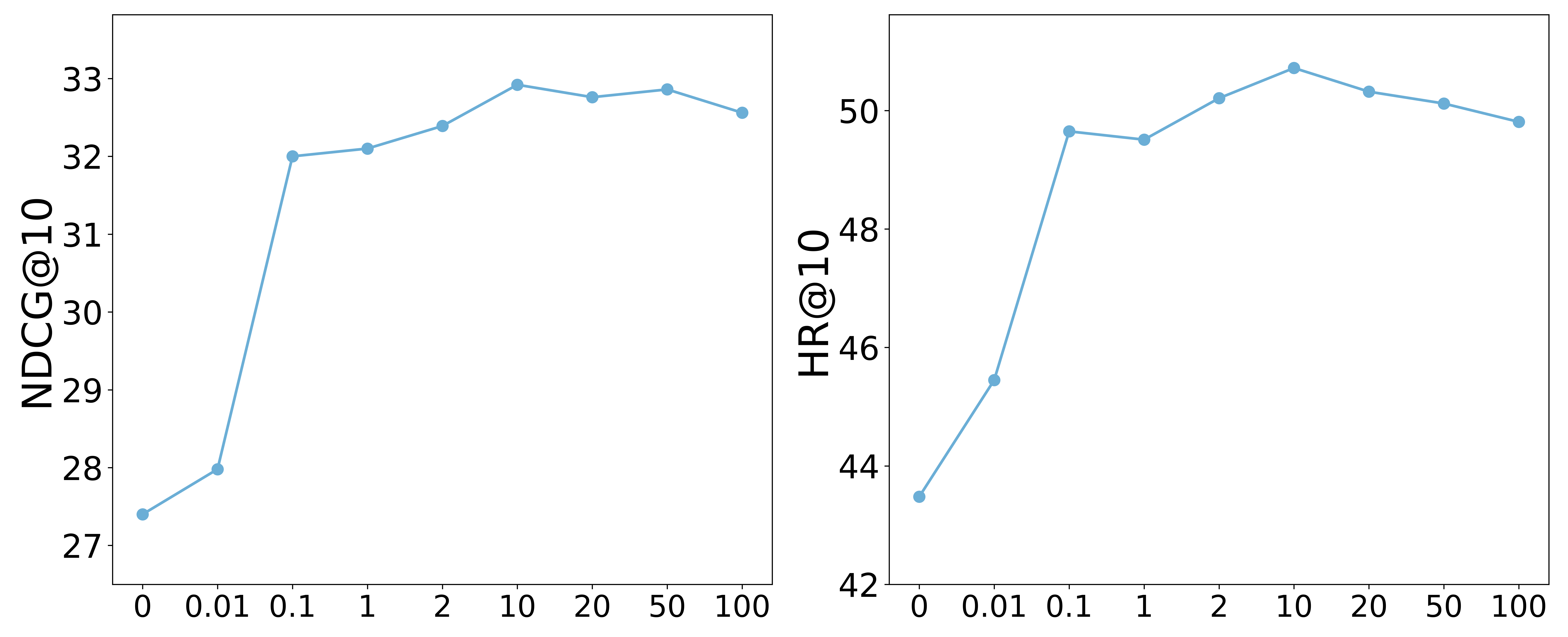}}

\caption{Results of different pre-training dataset sizes. The horizontal axis represents the proportion of the original pre-training dataset (\%).}
\label{fig:dataset}
\end{figure}

\subsection{Analysis on Pre-training Sizes.}

The Pre-training has been verified to benefit RECFORMER in SR on new domains. In Table \ref{tab:main}, our PT setting also achieves satisfactory results. In this section, we explore the influence of pre-training dataset size on the effectiveness of our initialization by conducting experiments with PLMs pre-trained on randomly selected subsets of 0.01\%, 0.1\%, 1\%, 2\%, 10\%, 20\%, and 50\% from the original pre-training dataset. From Fig. \ref{fig:dataset}, we can observe that leveraging pre-training on behavioral datasets could significantly improve our performance.
It is also astonishing that even a small amount of behavior-based pre-training (e.g., involving several thousand users) could successfully add the behavior-level preference to PLMs and achieve large improvement compared to the original Longformer.
This observation reinforces the conclusion that the effectiveness of our method is primarily attributed to behavior-aware PLMs rather than the original PLMs. It also suggests that a minimal amount of pre-training behavioral data is possibly sufficient and transferable for new-domain SR tasks, implying that our method is practical and efficient based on a unified behavior-tuned PLM-based model.

\section{CONCLUSION and FUTURE WORK}

In this work, we deeply explore the effect of PLMs in SR, identifying significant model underutilization and redundancy of PLM in behavior sequence modeling by attention analyses and freezing/simplifying the sequence modeling of PLM. 
Next, we conduct extensive experiments and analyses on different PLM-based SR variants, discovering that the simplified sequence modeling of conventional ID-based SR models enhanced with the proposed behavior-tuned PLMs initialized item embeddings achieves a giant performance boost, while vanilla PLMs initializations are unable to obtain any improvements. In the future, we will explore the scalability of both SR datasets (including the size and domain diversity) and PLMs, determining whether any scaling laws apply to our proposed initialization. Additionally, identifying more appropriate training objectives to align PLMs with behavioral knowledge for initialization presents a promising avenue for research.
\begin{acks}
This work is supported by the Young Elite Scientists Sponsorship Program by CAST (2023QNRC001)
\end{acks}

\bibliographystyle{ACM-Reference-Format}
\bibliography{sample-base}


\begin{thebibliography}{50}


\ifx \showCODEN    \undefined \def \showCODEN     #1{\unskip}     \fi
\ifx \showDOI      \undefined \def \showDOI       #1{#1}\fi
\ifx \showISBNx    \undefined \def \showISBNx     #1{\unskip}     \fi
\ifx \showISBNxiii \undefined \def \showISBNxiii  #1{\unskip}     \fi
\ifx \showISSN     \undefined \def \showISSN      #1{\unskip}     \fi
\ifx \showLCCN     \undefined \def \showLCCN      #1{\unskip}     \fi
\ifx \shownote     \undefined \def \shownote      #1{#1}          \fi
\ifx \showarticletitle \undefined \def \showarticletitle #1{#1}   \fi
\ifx \showURL      \undefined \def \showURL       {\relax}        \fi
\providecommand\bibfield[2]{#2}
\providecommand\bibinfo[2]{#2}
\providecommand\natexlab[1]{#1}
\providecommand\showeprint[2][]{arXiv:#2}

\bibitem[Achiam et~al\mbox{.}(2023)]%
        {gpt4}
\bibfield{author}{\bibinfo{person}{Josh Achiam}, \bibinfo{person}{Steven Adler}, \bibinfo{person}{Sandhini Agarwal}, \bibinfo{person}{Lama Ahmad}, \bibinfo{person}{Ilge Akkaya}, \bibinfo{person}{Florencia~Leoni Aleman}, \bibinfo{person}{Diogo Almeida}, \bibinfo{person}{Janko Altenschmidt}, \bibinfo{person}{Sam Altman}, \bibinfo{person}{Shyamal Anadkat}, {et~al\mbox{.}}} \bibinfo{year}{2023}\natexlab{}.
\newblock \showarticletitle{Gpt-4 technical report}.
\newblock \bibinfo{journal}{\emph{arXiv preprint arXiv:2303.08774}} (\bibinfo{year}{2023}).
\newblock


\bibitem[Bao et~al\mbox{.}(2023a)]%
        {bao2023bi}
\bibfield{author}{\bibinfo{person}{Keqin Bao}, \bibinfo{person}{Jizhi Zhang}, \bibinfo{person}{Wenjie Wang}, \bibinfo{person}{Yang Zhang}, \bibinfo{person}{Zhengyi Yang}, \bibinfo{person}{Yancheng Luo}, \bibinfo{person}{Fuli Feng}, \bibinfo{person}{Xiangnaan He}, {and} \bibinfo{person}{Qi Tian}.} \bibinfo{year}{2023}\natexlab{a}.
\newblock \showarticletitle{A bi-step grounding paradigm for large language models in recommendation systems}.
\newblock \bibinfo{journal}{\emph{arXiv preprint arXiv:2308.08434}} (\bibinfo{year}{2023}).
\newblock


\bibitem[Bao et~al\mbox{.}(2023b)]%
        {bao2023tallrec}
\bibfield{author}{\bibinfo{person}{Keqin Bao}, \bibinfo{person}{Jizhi Zhang}, \bibinfo{person}{Yang Zhang}, \bibinfo{person}{Wenjie Wang}, \bibinfo{person}{Fuli Feng}, {and} \bibinfo{person}{Xiangnan He}.} \bibinfo{year}{2023}\natexlab{b}.
\newblock \showarticletitle{TALLRec: An Effective and Efficient Tuning Framework to Align Large Language Model with Recommendation}.
\newblock \bibinfo{journal}{\emph{arXiv preprint arXiv:2305.00447}} (\bibinfo{year}{2023}).
\newblock


\bibitem[Beltagy et~al\mbox{.}(2020)]%
        {longformer}
\bibfield{author}{\bibinfo{person}{Iz Beltagy}, \bibinfo{person}{Matthew~E Peters}, {and} \bibinfo{person}{Arman Cohan}.} \bibinfo{year}{2020}\natexlab{}.
\newblock \showarticletitle{Longformer: The long-document transformer}.
\newblock \bibinfo{journal}{\emph{arXiv preprint arXiv:2004.05150}} (\bibinfo{year}{2020}).
\newblock


\bibitem[Brown et~al\mbox{.}(2020)]%
        {GPT3}
\bibfield{author}{\bibinfo{person}{Tom Brown}, \bibinfo{person}{Benjamin Mann}, \bibinfo{person}{Nick Ryder}, \bibinfo{person}{Melanie Subbiah}, \bibinfo{person}{Jared~D Kaplan}, \bibinfo{person}{Prafulla Dhariwal}, \bibinfo{person}{Arvind Neelakantan}, \bibinfo{person}{Pranav Shyam}, \bibinfo{person}{Girish Sastry}, \bibinfo{person}{Amanda Askell}, {et~al\mbox{.}}} \bibinfo{year}{2020}\natexlab{}.
\newblock \showarticletitle{Language models are few-shot learners}.
\newblock \bibinfo{journal}{\emph{Advances in neural information processing systems}}  \bibinfo{volume}{33} (\bibinfo{year}{2020}), \bibinfo{pages}{1877--1901}.
\newblock


\bibitem[Cao et~al\mbox{.}(2022)]%
        {cao2022contrastive}
\bibfield{author}{\bibinfo{person}{Jiangxia Cao}, \bibinfo{person}{Xin Cong}, \bibinfo{person}{Jiawei Sheng}, \bibinfo{person}{Tingwen Liu}, {and} \bibinfo{person}{Bin Wang}.} \bibinfo{year}{2022}\natexlab{}.
\newblock \showarticletitle{Contrastive Cross-Domain Sequential Recommendation}. In \bibinfo{booktitle}{\emph{Proceedings of the 31st ACM International Conference on Information \& Knowledge Management}}. \bibinfo{pages}{138--147}.
\newblock


\bibitem[Chang et~al\mbox{.}(2021)]%
        {gnnsr}
\bibfield{author}{\bibinfo{person}{Jianxin Chang}, \bibinfo{person}{Chen Gao}, \bibinfo{person}{Yu Zheng}, \bibinfo{person}{Yiqun Hui}, \bibinfo{person}{Yanan Niu}, \bibinfo{person}{Yang Song}, \bibinfo{person}{Depeng Jin}, {and} \bibinfo{person}{Yong Li}.} \bibinfo{year}{2021}\natexlab{}.
\newblock \showarticletitle{Sequential recommendation with graph neural networks}. In \bibinfo{booktitle}{\emph{Proceedings of the 44th international ACM SIGIR conference on research and development in information retrieval}}. \bibinfo{pages}{378--387}.
\newblock


\bibitem[Chen et~al\mbox{.}(2022)]%
        {chen2022intent}
\bibfield{author}{\bibinfo{person}{Yongjun Chen}, \bibinfo{person}{Zhiwei Liu}, \bibinfo{person}{Jia Li}, \bibinfo{person}{Julian McAuley}, {and} \bibinfo{person}{Caiming Xiong}.} \bibinfo{year}{2022}\natexlab{}.
\newblock \showarticletitle{Intent contrastive learning for sequential recommendation}. In \bibinfo{booktitle}{\emph{Proceedings of the ACM Web Conference 2022}}. \bibinfo{pages}{2172--2182}.
\newblock


\bibitem[Cho et~al\mbox{.}(2014)]%
        {GRU}
\bibfield{author}{\bibinfo{person}{Kyunghyun Cho}, \bibinfo{person}{Bart Van~Merri{\"e}nboer}, \bibinfo{person}{Caglar Gulcehre}, \bibinfo{person}{Dzmitry Bahdanau}, \bibinfo{person}{Fethi Bougares}, \bibinfo{person}{Holger Schwenk}, {and} \bibinfo{person}{Yoshua Bengio}.} \bibinfo{year}{2014}\natexlab{}.
\newblock \showarticletitle{Learning phrase representations using RNN encoder-decoder for statistical machine translation}.
\newblock \bibinfo{journal}{\emph{arXiv preprint arXiv:1406.1078}} (\bibinfo{year}{2014}).
\newblock


\bibitem[Cui et~al\mbox{.}(2022a)]%
        {cui2022m6}
\bibfield{author}{\bibinfo{person}{Zeyu Cui}, \bibinfo{person}{Jianxin Ma}, \bibinfo{person}{Chang Zhou}, \bibinfo{person}{Jingren Zhou}, {and} \bibinfo{person}{Hongxia Yang}.} \bibinfo{year}{2022}\natexlab{a}.
\newblock \showarticletitle{M6-Rec: Generative Pretrained Language Models are Open-Ended Recommender Systems}.
\newblock \bibinfo{journal}{\emph{arXiv preprint arXiv:2205.08084}} (\bibinfo{year}{2022}).
\newblock


\bibitem[Cui et~al\mbox{.}(2022b)]%
        {M6-rec}
\bibfield{author}{\bibinfo{person}{Zeyu Cui}, \bibinfo{person}{Jianxin Ma}, \bibinfo{person}{Chang Zhou}, \bibinfo{person}{Jingren Zhou}, {and} \bibinfo{person}{Hongxia Yang}.} \bibinfo{year}{2022}\natexlab{b}.
\newblock \showarticletitle{M6-Rec: Generative Pretrained Language Models are Open-Ended Recommender Systems}.
\newblock \bibinfo{journal}{\emph{arXiv preprint arXiv:2205.08084}} (\bibinfo{year}{2022}).
\newblock


\bibitem[Dallmann et~al\mbox{.}(2021)]%
        {metric2}
\bibfield{author}{\bibinfo{person}{Alexander Dallmann}, \bibinfo{person}{Daniel Zoller}, {and} \bibinfo{person}{Andreas Hotho}.} \bibinfo{year}{2021}\natexlab{}.
\newblock \showarticletitle{A case study on sampling strategies for evaluating neural sequential item recommendation models}. In \bibinfo{booktitle}{\emph{Proceedings of the 15th ACM Conference on Recommender Systems}}. \bibinfo{pages}{505--514}.
\newblock


\bibitem[Donkers et~al\mbox{.}(2017)]%
        {Donkers_Loepp_Ziegler_2017}
\bibfield{author}{\bibinfo{person}{Tim Donkers}, \bibinfo{person}{Benedikt Loepp}, {and} \bibinfo{person}{J{\"u}rgen Ziegler}.} \bibinfo{year}{2017}\natexlab{}.
\newblock \showarticletitle{Sequential user-based recurrent neural network recommendations}. In \bibinfo{booktitle}{\emph{Proceedings of the eleventh ACM conference on recommender systems}}. \bibinfo{pages}{152--160}.
\newblock


\bibitem[Gao et~al\mbox{.}(2023)]%
        {gao2023chat}
\bibfield{author}{\bibinfo{person}{Yunfan Gao}, \bibinfo{person}{Tao Sheng}, \bibinfo{person}{Youlin Xiang}, \bibinfo{person}{Yun Xiong}, \bibinfo{person}{Haofen Wang}, {and} \bibinfo{person}{Jiawei Zhang}.} \bibinfo{year}{2023}\natexlab{}.
\newblock \showarticletitle{Chat-rec: Towards interactive and explainable llms-augmented recommender system}.
\newblock \bibinfo{journal}{\emph{arXiv preprint arXiv:2303.14524}} (\bibinfo{year}{2023}).
\newblock


\bibitem[Geng et~al\mbox{.}(2022)]%
        {geng2022recommendation}
\bibfield{author}{\bibinfo{person}{Shijie Geng}, \bibinfo{person}{Shuchang Liu}, \bibinfo{person}{Zuohui Fu}, \bibinfo{person}{Yingqiang Ge}, {and} \bibinfo{person}{Yongfeng Zhang}.} \bibinfo{year}{2022}\natexlab{}.
\newblock \showarticletitle{Recommendation as language processing (rlp): A unified pretrain, personalized prompt \& predict paradigm (p5)}. In \bibinfo{booktitle}{\emph{Proceedings of the 16th ACM Conference on Recommender Systems}}. \bibinfo{pages}{299--315}.
\newblock


\bibitem[Guo et~al\mbox{.}(2022)]%
        {gnn2022evolutionary}
\bibfield{author}{\bibinfo{person}{Jiayan Guo}, \bibinfo{person}{Peiyan Zhang}, \bibinfo{person}{Chaozhuo Li}, \bibinfo{person}{Xing Xie}, \bibinfo{person}{Yan Zhang}, {and} \bibinfo{person}{Sunghun Kim}.} \bibinfo{year}{2022}\natexlab{}.
\newblock \showarticletitle{Evolutionary preference learning via graph nested gru ode for session-based recommendation}. In \bibinfo{booktitle}{\emph{Proceedings of the 31st ACM international conference on information \& knowledge management}}. \bibinfo{pages}{624--634}.
\newblock


\bibitem[Hao et~al\mbox{.}(2021)]%
        {hao2021adversarial}
\bibfield{author}{\bibinfo{person}{Xiaobo Hao}, \bibinfo{person}{Yudan Liu}, \bibinfo{person}{Ruobing Xie}, \bibinfo{person}{Kaikai Ge}, \bibinfo{person}{Linyao Tang}, \bibinfo{person}{Xu Zhang}, {and} \bibinfo{person}{Leyu Lin}.} \bibinfo{year}{2021}\natexlab{}.
\newblock \showarticletitle{Adversarial feature translation for multi-domain recommendation}. In \bibinfo{booktitle}{\emph{Proceedings of the 27th ACM SIGKDD Conference on Knowledge Discovery \& Data Mining}}. \bibinfo{pages}{2964--2973}.
\newblock


\bibitem[Harte et~al\mbox{.}(2023)]%
        {LLM-SR}
\bibfield{author}{\bibinfo{person}{Jesse Harte}, \bibinfo{person}{Wouter Zorgdrager}, \bibinfo{person}{Panos Louridas}, \bibinfo{person}{Asterios Katsifodimos}, \bibinfo{person}{Dietmar Jannach}, {and} \bibinfo{person}{Marios Fragkoulis}.} \bibinfo{year}{2023}\natexlab{}.
\newblock \showarticletitle{Leveraging large language models for sequential recommendation}. In \bibinfo{booktitle}{\emph{Proceedings of the 17th ACM Conference on Recommender Systems}}. \bibinfo{pages}{1096--1102}.
\newblock


\bibitem[He et~al\mbox{.}(2017)]%
        {He_Kang_McAuley_2017}
\bibfield{author}{\bibinfo{person}{Ruining He}, \bibinfo{person}{Wang-Cheng Kang}, {and} \bibinfo{person}{Julian McAuley}.} \bibinfo{year}{2017}\natexlab{}.
\newblock \showarticletitle{Translation-based recommendation}. In \bibinfo{booktitle}{\emph{Proceedings of the eleventh ACM conference on recommender systems}}. \bibinfo{pages}{161--169}.
\newblock


\bibitem[He and McAuley(2016)]%
        {Ruining_Julian_2016}
\bibfield{author}{\bibinfo{person}{Ruining He} {and} \bibinfo{person}{Julian McAuley}.} \bibinfo{year}{2016}\natexlab{}.
\newblock \showarticletitle{Fusing similarity models with markov chains for sparse sequential recommendation}. In \bibinfo{booktitle}{\emph{2016 IEEE 16th international conference on data mining (ICDM)}}. IEEE, \bibinfo{pages}{191--200}.
\newblock


\bibitem[Hidasi et~al\mbox{.}(2015)]%
        {GRURec}
\bibfield{author}{\bibinfo{person}{Bal{\'a}zs Hidasi}, \bibinfo{person}{Alexandros Karatzoglou}, \bibinfo{person}{Linas Baltrunas}, {and} \bibinfo{person}{Domonkos Tikk}.} \bibinfo{year}{2015}\natexlab{}.
\newblock \showarticletitle{Session-based recommendations with recurrent neural networks}.
\newblock \bibinfo{journal}{\emph{arXiv preprint arXiv:1511.06939}} (\bibinfo{year}{2015}).
\newblock


\bibitem[Hou et~al\mbox{.}(2022)]%
        {UniRec}
\bibfield{author}{\bibinfo{person}{Yupeng Hou}, \bibinfo{person}{Shanlei Mu}, \bibinfo{person}{Wayne~Xin Zhao}, \bibinfo{person}{Yaliang Li}, \bibinfo{person}{Bolin Ding}, {and} \bibinfo{person}{Ji-Rong Wen}.} \bibinfo{year}{2022}\natexlab{}.
\newblock \showarticletitle{Towards Universal Sequence Representation Learning for Recommender Systems}. In \bibinfo{booktitle}{\emph{Proceedings of the 28th ACM SIGKDD Conference on Knowledge Discovery and Data Mining}}. \bibinfo{pages}{585--593}.
\newblock


\bibitem[Hou et~al\mbox{.}(2023)]%
        {hou2023large}
\bibfield{author}{\bibinfo{person}{Yupeng Hou}, \bibinfo{person}{Junjie Zhang}, \bibinfo{person}{Zihan Lin}, \bibinfo{person}{Hongyu Lu}, \bibinfo{person}{Ruobing Xie}, \bibinfo{person}{Julian McAuley}, {and} \bibinfo{person}{Wayne~Xin Zhao}.} \bibinfo{year}{2023}\natexlab{}.
\newblock \showarticletitle{Large language models are zero-shot rankers for recommender systems}.
\newblock \bibinfo{journal}{\emph{arXiv preprint arXiv:2305.08845}} (\bibinfo{year}{2023}).
\newblock


\bibitem[Kang and McAuley(2018)]%
        {SASRec}
\bibfield{author}{\bibinfo{person}{Wang-Cheng Kang} {and} \bibinfo{person}{Julian McAuley}.} \bibinfo{year}{2018}\natexlab{}.
\newblock \showarticletitle{Self-attentive sequential recommendation}. In \bibinfo{booktitle}{\emph{2018 IEEE international conference on data mining (ICDM)}}. IEEE, \bibinfo{pages}{197--206}.
\newblock


\bibitem[Kang et~al\mbox{.}(2023)]%
        {kang2023llms}
\bibfield{author}{\bibinfo{person}{Wang-Cheng Kang}, \bibinfo{person}{Jianmo Ni}, \bibinfo{person}{Nikhil Mehta}, \bibinfo{person}{Maheswaran Sathiamoorthy}, \bibinfo{person}{Lichan Hong}, \bibinfo{person}{Ed Chi}, {and} \bibinfo{person}{Derek~Zhiyuan Cheng}.} \bibinfo{year}{2023}\natexlab{}.
\newblock \showarticletitle{Do LLMs Understand User Preferences? Evaluating LLMs On User Rating Prediction}.
\newblock \bibinfo{journal}{\emph{arXiv preprint arXiv:2305.06474}} (\bibinfo{year}{2023}).
\newblock


\bibitem[Krichene and Rendle(2020)]%
        {metric}
\bibfield{author}{\bibinfo{person}{Walid Krichene} {and} \bibinfo{person}{Steffen Rendle}.} \bibinfo{year}{2020}\natexlab{}.
\newblock \showarticletitle{On sampled metrics for item recommendation}. In \bibinfo{booktitle}{\emph{Proceedings of the 26th ACM SIGKDD international conference on knowledge discovery \& data mining}}. \bibinfo{pages}{1748--1757}.
\newblock


\bibitem[Li et~al\mbox{.}(2023)]%
        {li2023text}
\bibfield{author}{\bibinfo{person}{Jiacheng Li}, \bibinfo{person}{Ming Wang}, \bibinfo{person}{Jin Li}, \bibinfo{person}{Jinmiao Fu}, \bibinfo{person}{Xin Shen}, \bibinfo{person}{Jingbo Shang}, {and} \bibinfo{person}{Julian McAuley}.} \bibinfo{year}{2023}\natexlab{}.
\newblock \showarticletitle{Text Is All You Need: Learning Language Representations for Sequential Recommendation}.
\newblock \bibinfo{journal}{\emph{arXiv preprint arXiv:2305.13731}} (\bibinfo{year}{2023}).
\newblock


\bibitem[Liu et~al\mbox{.}(2023)]%
        {liu2023chatgpt}
\bibfield{author}{\bibinfo{person}{Junling Liu}, \bibinfo{person}{Chao Liu}, \bibinfo{person}{Renjie Lv}, \bibinfo{person}{Kang Zhou}, {and} \bibinfo{person}{Yan Zhang}.} \bibinfo{year}{2023}\natexlab{}.
\newblock \showarticletitle{Is ChatGPT a Good Recommender? A Preliminary Study}.
\newblock \bibinfo{journal}{\emph{arXiv preprint arXiv:2304.10149}} (\bibinfo{year}{2023}).
\newblock


\bibitem[Ma et~al\mbox{.}(2019)]%
        {ma2019pi}
\bibfield{author}{\bibinfo{person}{Muyang Ma}, \bibinfo{person}{Pengjie Ren}, \bibinfo{person}{Yujie Lin}, \bibinfo{person}{Zhumin Chen}, \bibinfo{person}{Jun Ma}, {and} \bibinfo{person}{Maarten~de Rijke}.} \bibinfo{year}{2019}\natexlab{}.
\newblock \showarticletitle{$\pi$-net: A parallel information-sharing network for shared-account cross-domain sequential recommendations}. In \bibinfo{booktitle}{\emph{Proceedings of the 42nd international ACM SIGIR conference on research and development in information retrieval}}. \bibinfo{pages}{685--694}.
\newblock


\bibitem[Mu et~al\mbox{.}(2023)]%
        {IDA-SR}
\bibfield{author}{\bibinfo{person}{Shanlei Mu}, \bibinfo{person}{Yupeng Hou}, \bibinfo{person}{Wayne~Xin Zhao}, \bibinfo{person}{Yaliang Li}, {and} \bibinfo{person}{Bolin Ding}.} \bibinfo{year}{2023}\natexlab{}.
\newblock \showarticletitle{ID-Agnostic User Behavior Pre-training for Sequential Recommendation}. In \bibinfo{booktitle}{\emph{Information Retrieval: 28th China Conference, CCIR 2022, Chongqing, China, September 16--18, 2022, Revised Selected Papers}}. Springer, \bibinfo{pages}{16--27}.
\newblock


\bibitem[Ni et~al\mbox{.}(2019)]%
        {amazon}
\bibfield{author}{\bibinfo{person}{Jianmo Ni}, \bibinfo{person}{Jiacheng Li}, {and} \bibinfo{person}{Julian McAuley}.} \bibinfo{year}{2019}\natexlab{}.
\newblock \showarticletitle{Justifying recommendations using distantly-labeled reviews and fine-grained aspects}. In \bibinfo{booktitle}{\emph{Proceedings of the 2019 conference on empirical methods in natural language processing and the 9th international joint conference on natural language processing (EMNLP-IJCNLP)}}. \bibinfo{pages}{188--197}.
\newblock


\bibitem[Ouyang et~al\mbox{.}(2022)]%
        {instruction-tuning}
\bibfield{author}{\bibinfo{person}{Long Ouyang}, \bibinfo{person}{Jeffrey Wu}, \bibinfo{person}{Xu Jiang}, \bibinfo{person}{Diogo Almeida}, \bibinfo{person}{Carroll Wainwright}, \bibinfo{person}{Pamela Mishkin}, \bibinfo{person}{Chong Zhang}, \bibinfo{person}{Sandhini Agarwal}, \bibinfo{person}{Katarina Slama}, \bibinfo{person}{Alex Ray}, {et~al\mbox{.}}} \bibinfo{year}{2022}\natexlab{}.
\newblock \showarticletitle{Training language models to follow instructions with human feedback}.
\newblock \bibinfo{journal}{\emph{Advances in neural information processing systems}}  \bibinfo{volume}{35} (\bibinfo{year}{2022}), \bibinfo{pages}{27730--27744}.
\newblock


\bibitem[Qiu et~al\mbox{.}(2022)]%
        {qiu2022contrastive}
\bibfield{author}{\bibinfo{person}{Ruihong Qiu}, \bibinfo{person}{Zi Huang}, \bibinfo{person}{Hongzhi Yin}, {and} \bibinfo{person}{Zijian Wang}.} \bibinfo{year}{2022}\natexlab{}.
\newblock \showarticletitle{Contrastive learning for representation degeneration problem in sequential recommendation}. In \bibinfo{booktitle}{\emph{Proceedings of the fifteenth ACM international conference on web search and data mining}}. \bibinfo{pages}{813--823}.
\newblock


\bibitem[Qu et~al\mbox{.}(2023)]%
        {qu2023thoroughly}
\bibfield{author}{\bibinfo{person}{Zekai Qu}, \bibinfo{person}{Ruobing Xie}, \bibinfo{person}{Chaojun Xiao}, \bibinfo{person}{Yuan Yao}, \bibinfo{person}{Zhiyuan Liu}, \bibinfo{person}{Fengzong Lian}, \bibinfo{person}{Zhanhui Kang}, {and} \bibinfo{person}{Jie Zhou}.} \bibinfo{year}{2023}\natexlab{}.
\newblock \showarticletitle{Thoroughly Modeling Multi-domain Pre-trained Recommendation as Language}.
\newblock \bibinfo{journal}{\emph{arXiv preprint arXiv:2310.13540}} (\bibinfo{year}{2023}).
\newblock


\bibitem[Ren et~al\mbox{.}(2023)]%
        {ren2023representation}
\bibfield{author}{\bibinfo{person}{Xubin Ren}, \bibinfo{person}{Wei Wei}, \bibinfo{person}{Lianghao Xia}, \bibinfo{person}{Lixin Su}, \bibinfo{person}{Suqi Cheng}, \bibinfo{person}{Junfeng Wang}, \bibinfo{person}{Dawei Yin}, {and} \bibinfo{person}{Chao Huang}.} \bibinfo{year}{2023}\natexlab{}.
\newblock \showarticletitle{Representation learning with large language models for recommendation}.
\newblock \bibinfo{journal}{\emph{arXiv preprint arXiv:2310.15950}} (\bibinfo{year}{2023}).
\newblock


\bibitem[Rendle et~al\mbox{.}(2010)]%
        {FPMC}
\bibfield{author}{\bibinfo{person}{Steffen Rendle}, \bibinfo{person}{Christoph Freudenthaler}, {and} \bibinfo{person}{Lars Schmidt-Thieme}.} \bibinfo{year}{2010}\natexlab{}.
\newblock \showarticletitle{Factorizing personalized markov chains for next-basket recommendation}. In \bibinfo{booktitle}{\emph{Proceedings of the 19th international conference on World wide web}}. \bibinfo{pages}{811--820}.
\newblock


\bibitem[Sun et~al\mbox{.}(2019)]%
        {BERT4Rec}
\bibfield{author}{\bibinfo{person}{Fei Sun}, \bibinfo{person}{Jun Liu}, \bibinfo{person}{Jian Wu}, \bibinfo{person}{Changhua Pei}, \bibinfo{person}{Xiao Lin}, \bibinfo{person}{Wenwu Ou}, {and} \bibinfo{person}{Peng Jiang}.} \bibinfo{year}{2019}\natexlab{}.
\newblock \showarticletitle{BERT4Rec: Sequential recommendation with bidirectional encoder representations from transformer}. In \bibinfo{booktitle}{\emph{Proceedings of the 28th ACM international conference on information and knowledge management}}. \bibinfo{pages}{1441--1450}.
\newblock


\bibitem[Tang and Wang(2018)]%
        {Tang_Wang_2018}
\bibfield{author}{\bibinfo{person}{Jiaxi Tang} {and} \bibinfo{person}{Ke Wang}.} \bibinfo{year}{2018}\natexlab{}.
\newblock \showarticletitle{Personalized top-n sequential recommendation via convolutional sequence embedding}. In \bibinfo{booktitle}{\emph{Proceedings of the eleventh ACM international conference on web search and data mining}}. \bibinfo{pages}{565--573}.
\newblock


\bibitem[Vaswani et~al\mbox{.}(2017)]%
        {transformers}
\bibfield{author}{\bibinfo{person}{Ashish Vaswani}, \bibinfo{person}{Noam Shazeer}, \bibinfo{person}{Niki Parmar}, \bibinfo{person}{Jakob Uszkoreit}, \bibinfo{person}{Llion Jones}, \bibinfo{person}{Aidan~N Gomez}, \bibinfo{person}{{\L}ukasz Kaiser}, {and} \bibinfo{person}{Illia Polosukhin}.} \bibinfo{year}{2017}\natexlab{}.
\newblock \showarticletitle{Attention is all you need}.
\newblock \bibinfo{journal}{\emph{Advances in neural information processing systems}}  \bibinfo{volume}{30} (\bibinfo{year}{2017}).
\newblock


\bibitem[Wang et~al\mbox{.}(2022b)]%
        {TransRec}
\bibfield{author}{\bibinfo{person}{Jie Wang}, \bibinfo{person}{Fajie Yuan}, \bibinfo{person}{Mingyue Cheng}, \bibinfo{person}{Joemon~M Jose}, \bibinfo{person}{Chenyun Yu}, \bibinfo{person}{Beibei Kong}, \bibinfo{person}{Zhijin Wang}, \bibinfo{person}{Bo Hu}, {and} \bibinfo{person}{Zang Li}.} \bibinfo{year}{2022}\natexlab{b}.
\newblock \showarticletitle{TransRec: Learning Transferable Recommendation from Mixture-of-Modality Feedback}.
\newblock \bibinfo{journal}{\emph{arXiv preprint}} (\bibinfo{year}{2022}).
\newblock


\bibitem[Wang and Lim(2023)]%
        {wang2023zero}
\bibfield{author}{\bibinfo{person}{Lei Wang} {and} \bibinfo{person}{Ee-Peng Lim}.} \bibinfo{year}{2023}\natexlab{}.
\newblock \showarticletitle{Zero-Shot Next-Item Recommendation using Large Pretrained Language Models}.
\newblock \bibinfo{journal}{\emph{arXiv preprint arXiv:2304.03153}} (\bibinfo{year}{2023}).
\newblock


\bibitem[Wang et~al\mbox{.}(2022a)]%
        {gnn2022exploiting}
\bibfield{author}{\bibinfo{person}{Nan Wang}, \bibinfo{person}{Shoujin Wang}, \bibinfo{person}{Yan Wang}, \bibinfo{person}{Quan~Z Sheng}, {and} \bibinfo{person}{Mehmet~A Orgun}.} \bibinfo{year}{2022}\natexlab{a}.
\newblock \showarticletitle{Exploiting intra-and inter-session dependencies for session-based recommendations}.
\newblock \bibinfo{journal}{\emph{World Wide Web}} \bibinfo{volume}{25}, \bibinfo{number}{1} (\bibinfo{year}{2022}), \bibinfo{pages}{425--443}.
\newblock


\bibitem[Wei et~al\mbox{.}(2023)]%
        {wei2023llmrec}
\bibfield{author}{\bibinfo{person}{Wei Wei}, \bibinfo{person}{Xubin Ren}, \bibinfo{person}{Jiabin Tang}, \bibinfo{person}{Qinyong Wang}, \bibinfo{person}{Lixin Su}, \bibinfo{person}{Suqi Cheng}, \bibinfo{person}{Junfeng Wang}, \bibinfo{person}{Dawei Yin}, {and} \bibinfo{person}{Chao Huang}.} \bibinfo{year}{2023}\natexlab{}.
\newblock \showarticletitle{Llmrec: Large language models with graph augmentation for recommendation}.
\newblock \bibinfo{journal}{\emph{arXiv preprint arXiv:2311.00423}} (\bibinfo{year}{2023}).
\newblock


\bibitem[Wu et~al\mbox{.}(2019)]%
        {SR-GNN}
\bibfield{author}{\bibinfo{person}{Shu Wu}, \bibinfo{person}{Yuyuan Tang}, \bibinfo{person}{Yanqiao Zhu}, \bibinfo{person}{Liang Wang}, \bibinfo{person}{Xing Xie}, {and} \bibinfo{person}{Tieniu Tan}.} \bibinfo{year}{2019}\natexlab{}.
\newblock \showarticletitle{Session-based recommendation with graph neural networks}. In \bibinfo{booktitle}{\emph{Proceedings of the AAAI conference on artificial intelligence}}, Vol.~\bibinfo{volume}{33}. \bibinfo{pages}{346--353}.
\newblock


\bibitem[Wu et~al\mbox{.}(2022)]%
        {wu2022selective}
\bibfield{author}{\bibinfo{person}{Yiqing Wu}, \bibinfo{person}{Ruobing Xie}, \bibinfo{person}{Yongchun Zhu}, \bibinfo{person}{Fuzhen Zhuang}, \bibinfo{person}{Ao Xiang}, \bibinfo{person}{Xu Zhang}, \bibinfo{person}{Leyu Lin}, {and} \bibinfo{person}{Qing He}.} \bibinfo{year}{2022}\natexlab{}.
\newblock \showarticletitle{Selective fairness in recommendation via prompts}. In \bibinfo{booktitle}{\emph{Proceedings of the 45th International ACM SIGIR Conference on Research and Development in Information Retrieval}}. \bibinfo{pages}{2657--2662}.
\newblock


\bibitem[Xie et~al\mbox{.}(2022)]%
        {xie2022contrastive}
\bibfield{author}{\bibinfo{person}{Xu Xie}, \bibinfo{person}{Fei Sun}, \bibinfo{person}{Zhaoyang Liu}, \bibinfo{person}{Shiwen Wu}, \bibinfo{person}{Jinyang Gao}, \bibinfo{person}{Jiandong Zhang}, \bibinfo{person}{Bolin Ding}, {and} \bibinfo{person}{Bin Cui}.} \bibinfo{year}{2022}\natexlab{}.
\newblock \showarticletitle{Contrastive learning for sequential recommendation}. In \bibinfo{booktitle}{\emph{2022 IEEE 38th international conference on data engineering (ICDE)}}. IEEE, \bibinfo{pages}{1259--1273}.
\newblock


\bibitem[Xu et~al\mbox{.}(2021)]%
        {LSSA}
\bibfield{author}{\bibinfo{person}{Chengfeng Xu}, \bibinfo{person}{Jian Feng}, \bibinfo{person}{Pengpeng Zhao}, \bibinfo{person}{Fuzhen Zhuang}, \bibinfo{person}{Deqing Wang}, \bibinfo{person}{Yanchi Liu}, {and} \bibinfo{person}{Victor~S Sheng}.} \bibinfo{year}{2021}\natexlab{}.
\newblock \showarticletitle{Long-and short-term self-attention network for sequential recommendation}.
\newblock \bibinfo{journal}{\emph{Neurocomputing}}  \bibinfo{volume}{423} (\bibinfo{year}{2021}), \bibinfo{pages}{580--589}.
\newblock


\bibitem[Yuan et~al\mbox{.}(2023)]%
        {yuan2023go}
\bibfield{author}{\bibinfo{person}{Zheng Yuan}, \bibinfo{person}{Fajie Yuan}, \bibinfo{person}{Yu Song}, \bibinfo{person}{Youhua Li}, \bibinfo{person}{Junchen Fu}, \bibinfo{person}{Fei Yang}, \bibinfo{person}{Yunzhu Pan}, {and} \bibinfo{person}{Yongxin Ni}.} \bibinfo{year}{2023}\natexlab{}.
\newblock \showarticletitle{Where to go next for recommender systems? id-vs. modality-based recommender models revisited}.
\newblock \bibinfo{journal}{\emph{arXiv preprint arXiv:2303.13835}} (\bibinfo{year}{2023}).
\newblock


\bibitem[Zhang et~al\mbox{.}(2023)]%
        {zhang2023recommendation}
\bibfield{author}{\bibinfo{person}{Junjie Zhang}, \bibinfo{person}{Ruobing Xie}, \bibinfo{person}{Yupeng Hou}, \bibinfo{person}{Wayne~Xin Zhao}, \bibinfo{person}{Leyu Lin}, {and} \bibinfo{person}{Ji-Rong Wen}.} \bibinfo{year}{2023}\natexlab{}.
\newblock \showarticletitle{Recommendation as instruction following: A large language model empowered recommendation approach}.
\newblock \bibinfo{journal}{\emph{arXiv preprint arXiv:2305.07001}} (\bibinfo{year}{2023}).
\newblock


\bibitem[Zhou et~al\mbox{.}(2020)]%
        {zhou2020s3}
\bibfield{author}{\bibinfo{person}{Kun Zhou}, \bibinfo{person}{Hui Wang}, \bibinfo{person}{Wayne~Xin Zhao}, \bibinfo{person}{Yutao Zhu}, \bibinfo{person}{Sirui Wang}, \bibinfo{person}{Fuzheng Zhang}, \bibinfo{person}{Zhongyuan Wang}, {and} \bibinfo{person}{Ji-Rong Wen}.} \bibinfo{year}{2020}\natexlab{}.
\newblock \showarticletitle{S3-rec: Self-supervised learning for sequential recommendation with mutual information maximization}. In \bibinfo{booktitle}{\emph{Proceedings of the 29th ACM international conference on information \& knowledge management}}. \bibinfo{pages}{1893--1902}.
\newblock


\end{thebibliography}

\end{document}